# Remarkable Disk and Off-nuclear Starburst Activity in the "Tadpole Galaxy" as revealed by the *Spitzer* Space Telescope

T.H. Jarrett[1], M. Polletta[2], I.P. Fournon[3], G. Stacey[4], K. Xu[1], B. Siana[2], D. Farrah[1], S. Berta[5], E. Hatziminaoglou[3], G. Rodighiero[5], J. Surace[1], D. Domingue[6], D. Shupe[1], F. Fang[1], C. Lonsdale[1], S. Oliver[7], M. Rowan-Robinson[8], G. Smith[2], T. Babbedge[8], E. Gonzalez-Solares[9], F. Masci[1], A. Franceschini[5], D. Padgett[1]

1. SSC/Caltech, 100-22 IPAC, Pasadena, CA 91125; 626-395-1844; jarrett@ipac.caltech.edu
2. University of California, San Diego, La Jolla, CA   92093-0424
3. Instituto de Astrofisica de Canarias, E38200 - La Laguna (Tenerife), Spain
4. Cornell University, 212 Space Sciences Building, Ithaca, NY 14853
5. Department of Astronomy, Padova University
6. Georgia College & State University, Dept of Chemistry and Physics, CBX 82, Milledgeville, GA 3106
7. Sussex University, Department of Physics and Astronomy, Brighton, East Sussex BN1 9QH
8. Imperial College London, Blackett Laboratory, Prince Consort Road, London, SW7 2AZ, UK
9. Cambridge University, Institute of Astronomy, Madingley Road, Cambridge. CB3 0HA, UK

## ABSTRACT

We present ground-based optical and *Spitzer* infrared imaging observations of the interacting galaxy UGC 10214, the "Tadpole Galaxy" (z = 0.0310), focusing on the star formation activity in the nuclear, disk, spiral arms and tidal tail regions. The ground-based optical data set spans a wavelength range between 0.3 to 0.8 μm, the NIR 1 - 2.2 μm, and the *Spitzer* infrared 3 to 70 μm. The major findings of this study are that the Tadpole is actively forming stars in the main disk outside of the nucleus and in the tidal plume, with an estimated mean star formation rate of ~2 to 4 $M_o$ yr$^{-1}$.  The most prominent sites of mid-infrared emission define a "ring" morphology that, combined with the overall morphology of the system, suggest the interaction may belong to the rare class of off-center collisional "ring" systems that form both shock-induced rings of star formation and tidal plumes. In stark contrast to the disk star formation, the nuclear emission is solely powered by older stars, with little evidence for ongoing star formation at the center of the Tadpole.  Extra-nuclear star formation accounts for >50% of the total star formation in the disk and spiral arms, featuring infrared-bright 'hot spots' that exhibit strong PAH emission , whose band strength is comparable to that of late-type star-forming disk galaxies.  The tidal tail, which extends 2′ (~75 kpc) into the intergalactic medium, is populated by super massive star clusters, M ~ $10^6$ $M_o$, likely triggered by the galaxy-galaxy interaction that has distorted UGC 10214 into its current "tadpole" shape. The Tadpole is therefore an example of an off-nuclear or tidal-tail starburst, with several large sites of massive star formation in the disk and in the plume, including the most prominent *HST*-revealed cluster, J160616.85+552640.6.  The clusters exhibit remarkable infrared (IR) properties, including exceptionally strong 24 μm emission relative to the underlying starlight, hot dust continuum, and PAH emission, with an estimated current star formation rate ~0.1 to 0.4 $M_o$ yr$^{-1}$, representing >10% of the total star formation in the system. We estimate the mass of the largest cluster to be ~1.4 - 1.6 × $10^6$ $M_o$ based on the g'-band (0.5 μm) and NIR (2.2 μm) integrated fluxes in combination with an assume M/L ratio appropriate to young clusters, or large enough to be classified as a nascent dwarf galaxy or globular cluster.

Subject headings: galaxies: individual (UGC10214; Arp 188; VV29) -- galaxies: ISM -- galaxies: interacting: infrared: galaxies -- stars: formation

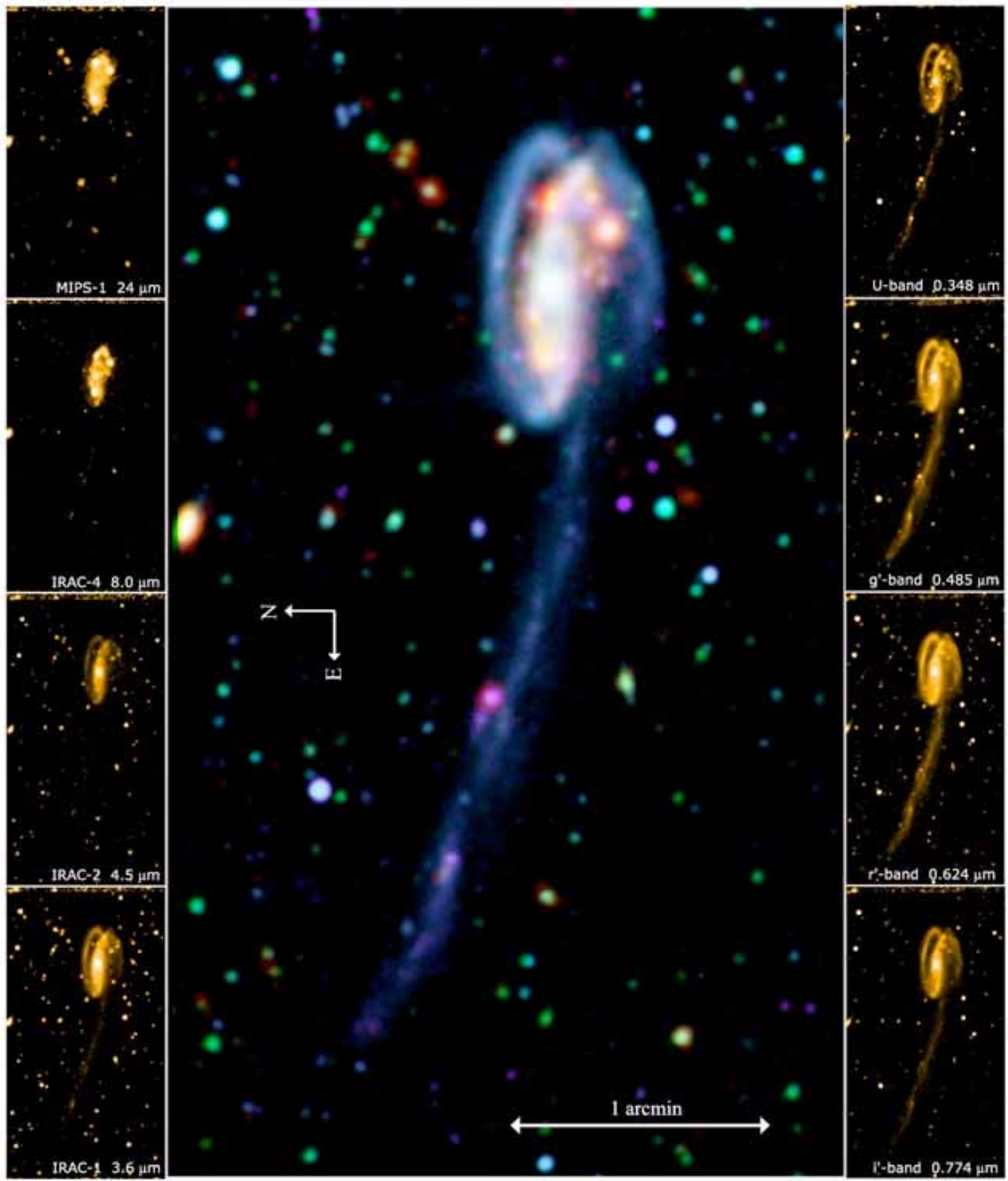

**Figure 1--** Optical-infrared panchromatic view of the Tadpole Galaxy (UGC10214). The center image is a composite combination of eight bands (from blue to red colors): U-band (0.348 μm), g'-band (0.485 μm), r'-band (0.624 μm), i'-band (0.774 μm), *Spitzer* IRAC-1 (3.6 μm), IRAC-2 (4.5 μm), IRAC-4 (8.0 μm) and MIPS-1 (24 μm). Maps of the individual bands are shown with side panels. The orientation is such that north is to the left and west to the top of the page. All images are shown with a modified n-σ log-sqrt stretch described in Jarrett et al. (2002).

## 1. Introduction

The Tadpole Galaxy (UGC10214 == VV 29 == Arp 188) comes by its colorful name from its distinctive shape that is the result of a close gravitational interaction with a smaller galaxy some ~100 to 200 Myr in the past (Briggs et al. 2001; de Grijs et al. 2003). Tidal forces have separated a major spiral arm from its main body, creating a long plume of gas and stars that extends over 2′ into the surrounding intergalactic medium, whose optical morphology is akin to the Toomre & Toomre (1972) classic weak galaxy-galaxy interaction tails. With a redshift of 0.0310 (Briggs et al. 2001), the Tadpole is located at a distance of 129 Mpc (adopting $H_o$ = 72 km/s/Mpc), implying that the main disk is ~38 kpc in projected size, and the tidal tail extends another 75 kpc from the disk. The disk exhibits a bar-like structure along the east-west axis, and is classified as 'SBc pec' Hubble Type in the NASA Extragalactic Database (NED). The main disk and the tidal tail are gas rich (Briggs et al. 2001), and thus providing a deep reservoir for star formation. IRAS FIR detections, F(IR) ≤ 1 Jy at 100 μm, indicate the presence of dust heated by massive star formation in the main disk. Recent high angular resolution *HST*-ACS optical imaging of the Tadpole system (Tran et al. 2003) and galaxies in the Tadpole field images (Benitez et al. 2004) reveal a spectacular network of newly forming, massive clusters that populate the spiral arms and tidal tail. Dozens of massive star clusters have been identified (de Grijs et al. 2003). A few of these clusters have earned the distinction of being "super star clusters", or super-massive star clusters, possessing total masses and stellar densities that rival globular clusters and dwarf galaxies ($M > 10^5 M_o$). The exact nature and future evolution of these super massive star clusters is still under lively debate, as studies concentrate on the connection between ongoing star formation in galaxies and proto-globular cluster formation in the environs of galaxy-galaxy interactions.

Galaxy-galaxy collisions and weaker tidal interactions, such as those inferred for the Tadpole, can boost the level of star formation activity within disk galaxies, in doing so transform the subsequent evolution of the merging system (Kennicutt et al. 1987; Lambas et al. 2003; Xilouris et al. 2004). Distortion of spiral arms and violent star formation episodes lead to large-scale alterations in the underlying gas and stellar populations. Under this intense compression, the interstellar medium (ISM) can rapidly coalesce with cloud-cloud collisions, accreting to form giant molecular clouds, creating the conditions for deeply embedded massive star formation (Olson & Kwan 1990; Scoville et al. 1991; Mihos & Hernquist 1996). One of the most spectacular examples is the Antennae Galaxy (NGC4038/9), revealed by the Infrared Space Observatory (ISO; Kessler et al. 1996), and recently, *Spitzer* Space Telescope, to be harboring off-nuclear embedded super massive clusters (Mirabel et al. 1998; Wang et al. 2004). In the outer spiral arms, atomic hydrogen is pulled and stretched into long filamentary tails, which will either dissipate into the intergalactic medium or eventually collapse back onto the parent galaxy or merger remnant forming streams or ring-like structures (Hibbard et al. 1994; Hibbard & Mihos 1995). Within these gas-rich tidal tails, super massive clusters are often found, signposts of the most vigorous and efficient star formation activity. In some cases the cluster density is high enough to create bound globular-cluster objects, or self-gravitating dwarf galaxies -- the so-called Tidal Dwarf Galaxy (e.g., Lisenfeld et al. 2002; 2004; Braine et al. 2001; Ho 1997; Holtzman et al. 1992; Kroupa 1998; Bastian et al 2005). Not all tidally interacting galaxies show remarkable star formation properties in comparison to field disk galaxies (e.g., Bergvall, Laurikainen & Aalto 2003), underscoring the complexity of triggered star formation and ensuing evolution of galaxies.

To properly understand the evolution of interacting galaxies, one must study the interaction between the underlying (old star) mass distribution, dust distribution, neutral atomic gas, cold molecular gas and newly forming (energetic) stars. Consequently, this sets a premium on building complete data sets that span the radio-ultraviolet electromagnetic spectrum. A rich data set was recently made available by the *Spitzer* Wide-field Infrared Extragalactic Survey (SWIRE; Lonsdale et al. 2003)[1]. Within the SWIRE field ELAIS N1 (Oliver et al. 2000) region lurks the Tadpole Galaxy and its associated group of neighboring galaxies. The data set includes deep optical imaging from 0.3 to 0.8 μm, near-infrared (NIR) imaging from 1 to 2 μm, *Spitzer* IRAC imaging from 3 to 9 μm, and *Spitzer* MIPS imaging at 24 and 70 μm, which may be combined with the existing radio and HST observations. The data sets are complementary with the optical and NIR measurements sensitive to young and old stars, respectively, while the mid-infrared (MIR) channels trace the interstellar dust emission and vibrational emission from polycyclic aromatic hydrocarbons (PAHs). Moreover, the high-galactic latitude ELAIS N1 field is ideally suited for extragalactic work due to minimal Galactic extinction and low confusion noise from foreground stars. This work combines the data sets to investigate the large-scale morphology of the disk and tidal tail, properties of the embedded star formation regions, and the super massive star clusters in the tail and spiral arms. The data and empirical model sets are described in §2, the global properties and detailed measurements in §3, and the star formation rates, SED, PAH strengths and color results are discussed in §4, followed by a summary of the major findings in §5.

## 2. Observations & Data Reduction

### 2.1 Optical Images

As part of the SWIRE project, deep optical imaging of the ELAIS N1 field was acquired in September 1999 with the digital Wide-field Camera aboard the 2.5-m Isaac Newton Telescope (INT) of the La Palma Observatory in the Canary Islands, Spain, as part of the INT Wide Field Survey (McMahon et al. 2001). UGC10214 and its environs are fully contained within large mosaics that were constructed from U-band (0.358 μm), g'-band (0.485 μm), r'-band (0.624 μm) and i'-band (0.774 μm) filtered images. The angular resolution is typically ~1″ (~625 pc), achieving a point source sensitivity ~24.2 (U), 25.5 (g'), 25.4 (r') and 23.6 (i') mag (S/N=5), calibrated on the Vega system. Flux densities were derived by converting to AB mags and applying the standard zero point scaling (3630.78 Jy). The photometric calibration uncertainty is better than 5%. All imaging data, including the optical mosaics, were reprojected and resampled onto a common grid and orientation of 0.2″ pixels with standard (E of N) orientation. For photometric comparison between the optical and IR images, the optical images were gaussian-convolved to match the angular resolution of the MIR images (see below). We also employ *HST*-ACS imaging (Tran et al. 2003) in *g* (F475W) and *V* (F606W) filters, which provide superior spatial resolution (~100 pc per beam) to detail the underlying young stellar population that dominates the total power coming from the Tadpole.

### 2.2 Optical Spectroscopy

Spectroscopy of the Tadpole nucleus and brightest tidal tail super star cluster, of which we give the designation J160616.85+552640.6, were acquired with the COSMIC camera mounted to the Hale 200" telescope of the Palomar Observatory in Southern California. The instrument was configured for long-slit grism spectroscopy using a 1.5″ slit width. The dispersion is ~3.1 Å per pixel, achieving a resolution of 8 Å per arcsec, with spectral coverage from 4000 to 9000 Å. The total exposure time for the Tadpole nucleus was 240 seconds, and 300 seconds for the super star cluster. The measurements were flux calibrated using spectroscopic calibration stars and validated via comparison with the corresponding broad-band fluxes. The observations were conducted during good photometric conditions, although the seeing was poor, ~3 to 4″, which incurs additional uncertainty (in addition to S/N) with the flux measurements. The 1-D spectra were extracted using a 15″ aperture. Residual telluric absorption bands are present in the extracted spectra (labeled A and B), while a 5892

---
[1] The SWIRE data presented in this paper are available through the Spitzer Science Center archive; http://ssc.spitzer.caltech.edu/

Å sodium-D emission line is also present in the extracted spectra (labeled "n.s." in Fig 5). The spectra (Figure 5) are presented in §3.2.

## 2.3 NIR Images

NIR J-band (1.2 μm) and Ks-band (2.2 μm) imaging was acquired in March 2003 with the Wide-Field Infrared Camera (Wilson et al. 2003) mounted to the Hale 200" telescope of the Palomar Observatory. The typical "seeing" was between 1 to 1.5″, with a point source sensitivity of J = 18.9 and Ks = 17.5 mag (S/N=5). The images were flux and astrometrically calibrated using 2MASS stellar photometry and coordinate positions, achieving ~7% rms uncertainty in the absolute flux calibration and sub-arcsecond astrometry, respectively. Flux densities were derived using the Vega-system of 2MASS (Cutri et al. 2000). A calibration check was carried out by comparing the integrated flux of the UGC10214 nucleus with that measured by 2MASS Large Galaxy Atlas (Jarrett et al. 2003). As with the optical imaging, the NIR images were reprojected/resampled onto a common grid, and were gaussian-convolved to match the MIR 8 μm imaging (see below).

## 2.4 MIR Images

Broadband IR images were acquired with the *Spitzer* Space Telescope using the Infrared Array Camera (IRAC; Fazio et al. 2004) and the Multiband Imaging Photometer (MIPS; Rieke et al. 2004). The four IRAC bands are: 3.6 μm, 4.5 μm, 5.8 μm and 8.0 μm, which have a corresponding angular resolution between 2.5 to 3″. MIPS includes three detectors, with only the 24 μm and 70 μm imaging results reported in this work. The angular resolution at 24 μm and 70 μm is ~6″ and 20″, respectively. The primary data reductions were carried out by the *Spitzer* Science Center (SSC) pipeline, version S10.5 (for the 70 μm maps, version S11.0 was used), creating astrometric and flux-calibrated images that were then combined into large mosaics using SWIRE mosaic tools (featuring SSC-developed MOPEX) that performed background matching and outlier (cosmic ray, latent artifact) rejection. For IRAC, point source depths of ~4 μJy to 40 μJy (S/N=5) are achieved for 3.6 to 8 μm respectively, with a relative photometric uncertainty ~4 to 5%, and absolute calibration uncertainty of ~10% under most condition. For MIPS 24 μm images, the sensitivity is ~150 μJy (S/N = 5) with absolute calibration estimated to be ~10%, and for 70 μm the sensitivity is ~20 mJy with a calibration uncertainty ~20%. Longer wavelength imaging was acquired at 160 μm, however the spatial resolution was not optimal for the Tadpole and is therefore not reported I this work. Additional details of the SWIRE project may be found in Lonsdale et al. (2003) and the data products in Surace et al. (2004). As with the optical and NIR imaging, the *Spitzer* IRAC data were reprojected and resampled onto a common grid, and were gaussian-convolved to match the 8μm images resolution.

## 2.5 Aperture Photometry

Detailed measurements are carried out using circular aperture photometry. The optimal aperture size is chosen under these considerations: angular scale size of the Tadpole components, spatial resolution of the *Spitzer* IRAC and MIPS observations, and the aperture size used to calibrate the standard Spitzer imaging products. Accordingly, we adopt a circular aperture with radius = 6″ (~3.8 kpc), matched at all wavelengths. This aperture is twice as small as the standard calibration aperture for *Spitzer* observations, which therefore requires the IRAC and MIPS photometry to be aperture-corrected using an empirically-derived curve-of-growth relation. This relation was derived by comparing the IR images with the appropriate detector point spread function, assuming the signal coming from the UGC10214 photometry is originating from point-like (unresolved) sources. This assumption is probably adequate for most of the measurements reported in this study, but systematic deviations become important for large angular scale structures that dominate the IR emission. Accordingly, we applied a modestly broad smoothing function to the images to minimize the systematic flux calibration offsets between resolved and unresolved sources. The appropriate aperture (flux scaling) corrections are as follows: 1.060, 1.063, 1.082, 1.092 and 1.563 corresponding to 3.6, 4.5, 5.8, 8.0 and 24.0 μm data. The 70 um imaging is only used to measure the global properties of the Tadpole using a large aperture that does not require an aperture correction. Finally, the MIPS data are further divided by 0.96 and 0.92, for 24 and 70 μm respectively, to correct the flux calibration from star-like (blue) emission to dusty spiral galaxy (red) emission, as recommended by the *Spitzer* Science Center MIPS user guidebook. Integrated fluxes are accompanied by an estimated uncertainty that includes both a formal error propagation (S/N) and inclusion of an error due to calibration and aperture correction uncertainty. Both IRAC and MIPS-24μm have calibration uncertainties that are expected to be ~10% (for small or standard calibration apertures). The MIPS-70 μm calibration is uncertain at the level of 20 to 30%.

## 2.6 Model SED Templates

Interpretation of the Tadpole spectral energy distribution (SED) is facilitated by comparison with model SED templates that include major stellar, dust and PAH components for representative types of galaxies that offer similarities and contrast with the Tadpole Galaxy disk and tidal tail. We refer to five different composite SEDs: (1) the underlying (old) stellar component is represented by the spectrum of 13 Gyr elliptical galaxies, which are (to first approximation) free of dust, (2) early-type disk galaxies are represented by an S0/Sa spiral, (3) more active star-forming disk galaxies are represented by the spectrum of an Sc disk spiral, (4) IR-bright galaxies are represented by the prototype starburst galaxy M82, characterized by a steeply rising SED from dust emission, and (5) active, dust absorbed nuclei by a Seyfert 2 template.

Except for the Seyfert 2 template, the SED templates were obtained using the GRASIL (GRAphite and SILicate) code (Silva et al. 1998). GRASIL generates galaxy SEDs from optical to far-infrared (FIR) wavelengths by computing the spectral evolution of stellar systems and taking into account the effects of dust in a variety of geometries. The galaxy templates used in this work were obtained using the parameters adopted in Silva et al. (1998). Since the current version of the code provides only an approximate treatment of the PAH features, we replaced the 5-20 μm spectral region, where PAH features dominate, with observed IR spectra of galaxies from the PHT-S spectrometer and ISO-SWS (Lu et al. 2003; Sturm et al. 2000) on the Infrared Space Observatory and from SINGS IRS data from the *Spitzer* Space Telescope. In one case we adjusted the intensity of the PAHs for an S0/Sa Type galaxy to better match the empirical SED of the Tadpole nucleus by scaling the IRS spectrum of NGC7331 from 5.4 to 25 μm (SINGS: Kennicutt et al. 2003; Smith et al. 2004). The model SED templates were normalized to the NIR 2.2 μm flux density measured for the Tadpole apertures (§2.5), thereby forcing the stellar contribution of the SEDs to match that of the Tadpole. The NIR represents the peak in stellar flux density for most galaxies (including the Tadpole) and is minimally effected by extinction, and is therefore an optimal window to match with the comparison SEDs. The Seyfert 2 model template was provided by one of the authors (Polletta 2005, in preparation).

## 3 Physical Content of Optical-IR Imaging

Spanning a sizeable portion of the electromagnetic spectrum, the SWIRE imaging provides lucid contrast between stellar populations and the nebular ISM that comprise the bulk of the Tadpole emission. The optical imaging separates the youngest, hottest stellar populations from the moderate aged (type A-F) stars. The NIR imaging illuminates the oldest, most numerous stellar populations that form the backbone of the disk and halo structures. The MIR, with $\lambda < 5$μm, combines the old (evolved) stellar population with a hot dust component originating from the ISM. The MIR, with $\lambda > 6$ μm, measures the warm dust continuum and molecular aromatic features (e.g., Leger & Puget 1984; Boulanger et al. 1998) that arise from the diffuse ISM and dense star formation regions. Figure 1 encapsulates the full informational content of the data set, assigning a separate color scale (from blue to red) to each band, optical to IR, respectively, to form the color composite.

The individual bands are shown on the perimeter with a monochromatic scale. The combined colors provide clues into the past and current state of the Tadpole, and combined with dynamical models, into the future evolution of the system. (The printed version of this figure does not do it full justice, the reader is urged to inspect the online full resolution version of this figure to gain maximum fidelity in details.) The lower resolution 70 µm *Spitzer* imaging is shown in Figure 2, overlaid with contours of 24 µm emission and HI column density (Briggs et al. 2001) to contrast the gas and dust in the system.

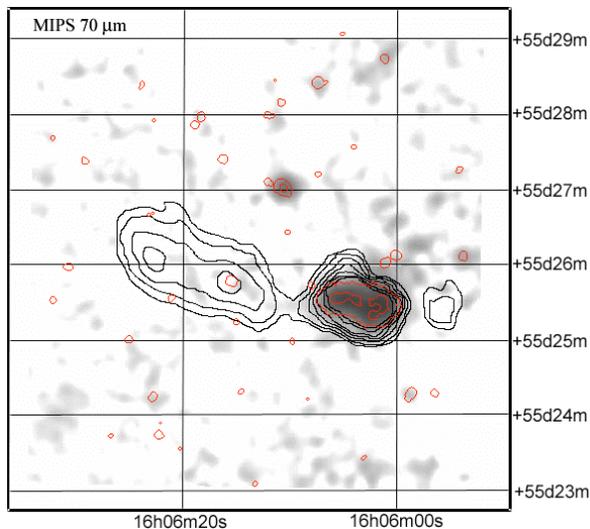

**Figure 2** -- *Spitzer*-**MIPS 70 µm map of the Tadpole Galaxy. The log grey-scale ranges from 0.2 MJy sr$^{-1}$ to the peak flux in the disk, ~5 MJy sr$^{-1}$. Overlaid in red are contours of the 24 µm map: 0.15 and 0.95 MJy sr$^{-1}$, and in black contours of HI column density: 0.6, 1.2, 2.4, 3.6, 4.8, 5.9 $\times 10^{20}$ H atoms cm$^{-2}$ (Briggs et al. 2001). The equatorial coordinates are in J2000.**

The most striking features seen in Figure 1 include: (a) the 'blue' spiral arms and tidal tail that are prominent in the bands that are sensitive to the youngest, most massive stars that are unobscured by dust, (b) IR-bright 'hot spots' in a ring-like structure that dominates the disk emission, indicative of embedded dusty star-formation regions, and (c) the prominent 24 µm 'hot spots' in the tidal plume that are spatially coincident with the brightest "super star clusters" resolved by *HST* (e.g., Tran et al. 2003;

De Grijs et al. 2003). The embedded star formation "ring" in the disk and spiral arms is most clearly visualized with comparison between the *HST* V-band emission and the *Spitzer* IRAC-8 µm emission; Figure 3. Dark dust lanes coincide with bright 8 µm emission, most notably along the eastern side of the Tadpole. Fainter structures, in Figures 1 and 3, can be discerned from the background, including a western plume, extending beyond the figure boundary, and a polar stream of stars that wraps under the disk before exiting northward along the eastern edge of the disk. In Figure1, the stream is horizontal, near the lower 1/3 of the disk. Figure 4 demarks the northern exit location of the stream. The gas dynamics of these features are discussed in Briggs et al. 2001. Interestingly, the 'bar' that is so prominent in optical photographs of the Tadpole may be an illusion of the disk inclination and distorted spiral arm structure, resembling more of a ring-like structure that connects arms that spiral and twist outward. The tidal interaction with the intruder galaxy, VV29c (in Figure 4, note the faint 'blue' fuzzy object hiding behind the western edge of the disk, at the extreme end of the 'bar') has not catastrophically disrupted the parent galaxy. On the other hand, the disk and tidal tail show evidence for recent starburst activity, both deeply embedded within dense molecular clouds (IR markers) and unobscured sites that are fully open to the disk and intergalactic medium -- as clearly shown with the *HST*-ACS imaging; note the optically-bright cluster in the disk, Figure 3, which coincides with an IR emission peak. The most massive star formation sites in the tidal tail have been postulated to be proto-globular clusters or even self-gravitating dwarf galaxies. In the following sections we will examine these exceptional regions in more detail, but first we present the global measurements for the Tadpole system.

## 3.1 Global Measurements of the System: Disk + Tidal Tail

Although global measurements represent a composite of the stellar and ISM components of galaxies, they are useful for comparison with measurements made of distant unresolved galaxies. Here we measure the total integrated light coming from the disk and tidal tail. This was accomplished by creating a "mask" that enveloped the system, eliminating both foreground (Milky Way) stars and background galaxies. Stars and galaxies are easily identified with colors; see for example Figure 1, where stars (bright and blue/white) and galaxies (typically 'green' due to strong 2.2 μm and 3.6 μm emission) are clearly discerned from Tadpole nebular emission. The end result is an integrated area of 2.97 sq. arcmin (equivalent to a 1.95 arcmin diameter circle). The results are given in Table 1, including the filter central wavelength, band-pass, flux density and monochromatic luminosity derived from $\nu L_\nu$. The formal uncertainties (S/N ratio) are small in principle, but the real limiting factor is the calibration systematics, which range from 5-10% for the optical and NIR measurements, and ~10 to 20% for the *Spitzer* measurements. The latter is driven by the uncertainty of expanding the standard point source calibration to extended sources. We have adopted IRAC flux correction values based on the work from Mike Pahre (2004, private communication; see also Pahre et al. 2004a), accordingly: multiply by 0.95, 0.95, 0.86 and 0.75, for IRAC-1, 2, 3 and 4, respectively[2]. Finally, we have not applied any reddening corrections to the integrated fluxes. We will consider extinction and reddening when we look at detailed regions, presented below.

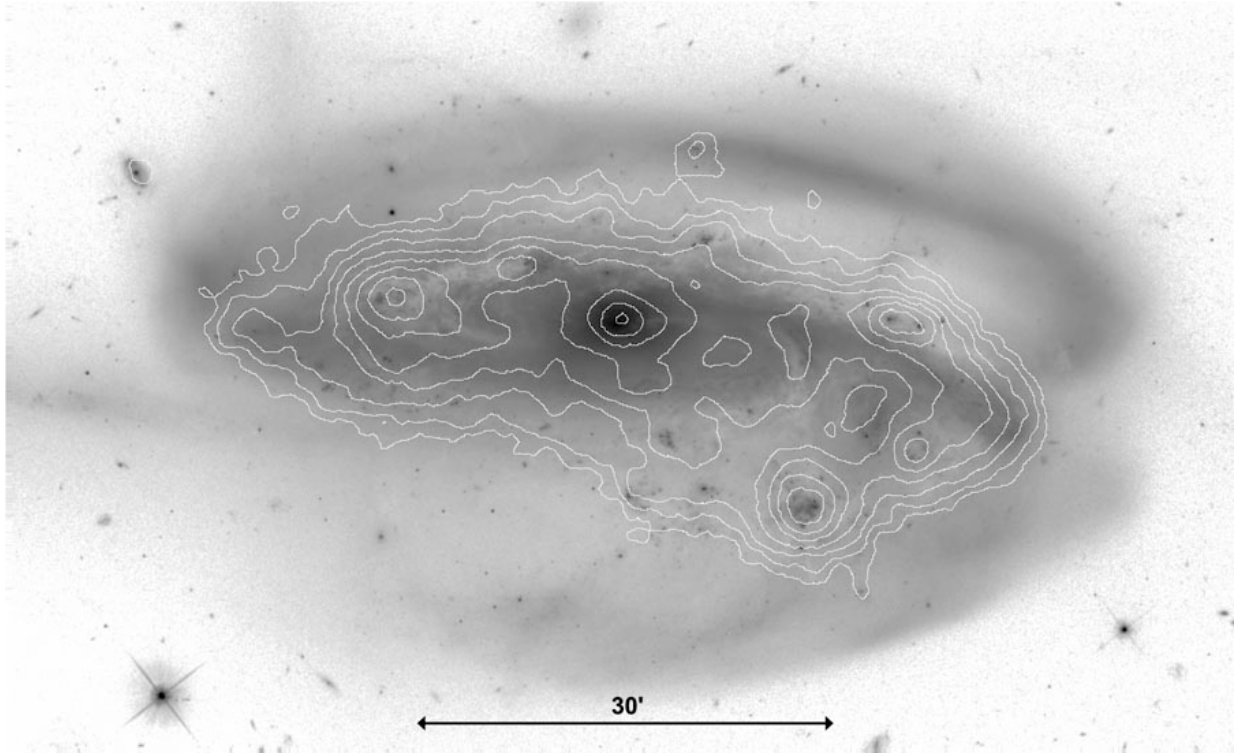

**Figure 3 -- *HST* and *Spitzer* views of the Tadpole disk. The grey-scale shows the V-band (F606W) from *HST*-ACS (Tran et al. 2003) at full spatial resolution. Overlaid are contours of *Spitzer* 8 μm imaging, ranging from 0.3 to 5 MJy sr$^{-1}$ in surface brightness. The orientation is standard, E of N.**

The global SED indicates a luminous Population I stellar component mixed with an underlying old star population. The global dust component, comparing the PAH emission at 8 μm to the dust continuum at 24 μm, has a ratio (unity) that is comparable to normal disk galaxies (e.g., Helou et al. 2004). The IRAS Point Source Catalog fluxes are: 0.12 ± 10%, 0.05±10%, 0.16 ± 20%, 0.78 ± 15% Jy at 12, 25, 60 and 100 μm respectively. The IRAS 25 and 60 μm measurements are in agreement (within the measurement uncertainties) with the *Spitzer* fluxes at 24 and 70 μm: 0.03 and 0.21 ± 20% Jy, respectively. For the luminosity in each band we assume a distance of 129 Mpc to the Tadpole system (see §1 for redshift information), and normalized by the Solar luminosity. The flux density peaks in the NIR, even as most of the energy is coming from the 0.6 to 1 μm optical/NIR window. The mid and FIR luminosity represents re-radiated photons from the hot, stellar component. Starlight is the principle component that powers the Tadpole Galaxy.

---

[2] The largest, and therefore most uncertain, correction is for the two longest wavelength IRAC channels.

## 3.2 Nucleus

Inspection of the monochromatic images in Figure 1 indicates a nucleus that has a strong stellar component relative to the ISM emission. Another view of the nucleus and disk of the Tadpole is given in Figure 4 (top panel), with an RGB color-combination between the *HST*-ASC g-band image (Tran et al. 2003), IRAC 3.6 μm, and MIPS 24 μm images. No attempt is made to match the disparate angular resolution between *HST* and *Spitzer*. The nucleus is unresolved at the distance of the Tadpole, with *HST* and *Spitzer* spatial resolutions of ~0.1 kpc and ~2 kpc, respectively. Nonetheless the stellar component is clearly dominating the total light output from the nucleus. Its composite color suggests a normal blend of aging stars, consistent with the individual global flux measurements. Extinction is present in the disk and nuclear regions, as dust lanes are clearly seen in the optical images (Figure 1; Figure 3). A quantitative investigation of the nucleus begins with the spectral energy distribution (SED) derived from our optical-IR data sets. The integrated light from the nuclear region is obtained using a 12″ diameter circular aperture centered on 16h06m03.9s +55d25m32s for each data set, corresponding to a physical size of 7.5 kpc. The aperture size was selected to optimally match with the 8 and 24 μm beams, while also minimizing the aperture correction required to conform to the standard Spitzer flux calibration.

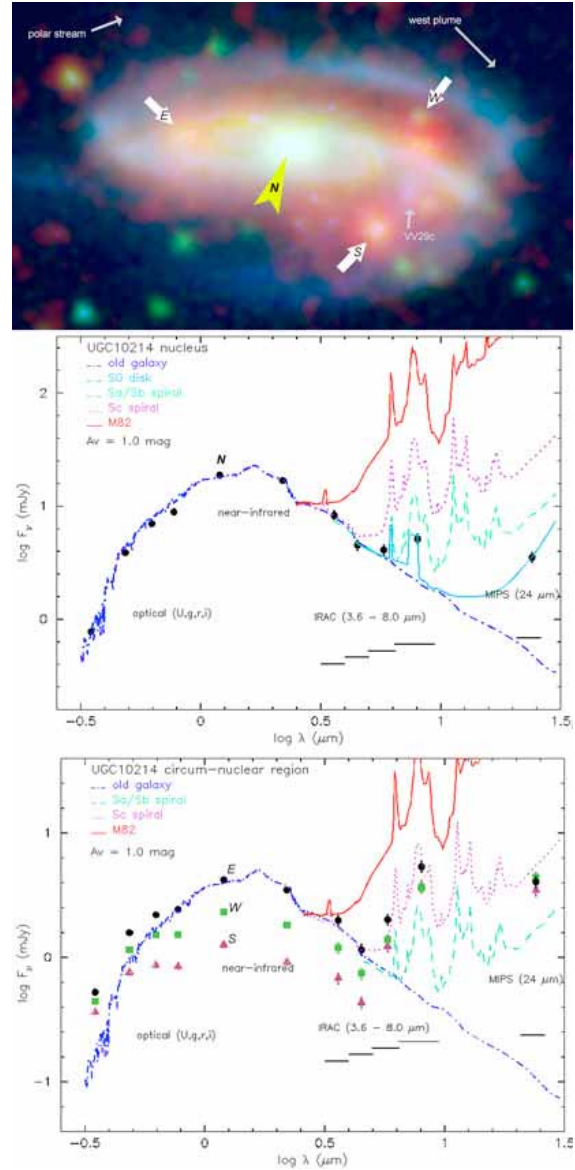

**Figure 4** -- **Tadpole Galaxy disk. The upper panel shows a 3-color RGB composite of the disk: blue is g-band (F475W) from *HST*-ACS (Tran et al. 2003), green is IRAC-1 (3.6 μm) and red is MIPS-1 (24 μm). The nucleus (*N*, yellow arrow) and three IR-bright regions are marked *E*, *W*, and *S*. The x-axis field of view is ~1′. The orientation is standard E of N. The middle panel shows the optical-infrared SED of the nucleus, integrated using a 12″ aperture. The lower panel shows SEDs for three different bright infrared regions in the disk (see top panel). For comparison, five empirical SED templates normalized to the 2.2 μm flux; adapted from Silva et al. 1998) are shown: blue corresponds to an old (dust free) L\* galaxy, cyan to an S0 disk; green is an Sa/Sb spiral, magenta is an Sc spiral, and red the integrated light from the dusty starburst M82.**

At 12″, a modest aperture correction is applied to the *Spitzer* integrated fluxes; see §2.5 for more details. The Tadpole SED, flux density versus wavelength, is given in Figure 4, middle panel, where each filter/band integrated flux is denoted with a filled circle and flux uncertainty error bar.

For comparison with the Tadpole nuclear measurements we employ five different empirical SED templates (described in §2.6). The comparison SEDs are normalized to the 2.2 µm flux density of the Tadpole. For clarity, we only show the old stellar component at short (optical) wavelengths, which is the basic shape for all the SEDs (there are of course emission features in the optical, but they are unimportant in this context). As shown in Figure 4, the Tadpole nuclear optical SED closely resembles that of an "old" dust-free galaxy. The model SEDs include a uniform visual extinction of ~1 mag applied to the Tadpole fluxes. For the red to NIR window we have combined the Cardelli et al. (1989) and Calzetti et al. (1996) reddening laws to compute the dust extinction (where the Calzetti reddening is more appropriate to dusty, starburst galaxies). For the MIR ($\lambda > 5$ µm) fluxes, we simply adopt $A_{\lambda>5\mu m} = 0.05$ Av, which should be adequate for most conditions unless silicate absorption at ~10 µm is present, in which case the extinction can easily double in this window. The derived extinction is only a crude 'global' estimate given that integrated light is coming from a relatively large region that includes non-uniform clumpy ISM. All the same, the spectrum of an old galaxy fits the Tadpole short wavelength light reasonably well -- the stellar component dominates the SED from 0.3 to 4.5 µm. The MIR shows a dust continuum signature with modestly elevated 8 and 24 µm emission, not unlike the early-type S0/Sa template shown for comparison, and is clearly not analogous to dust-rich or starburst galaxies. The slight IR excess probably arises from intermediate-age and evolved AGB stars. For completeness SED templates from AGN, LINER and Seyfert-like galaxies (§2.6) were compared with the Tadpole nuclear SED, showing no correlation whatsoever.

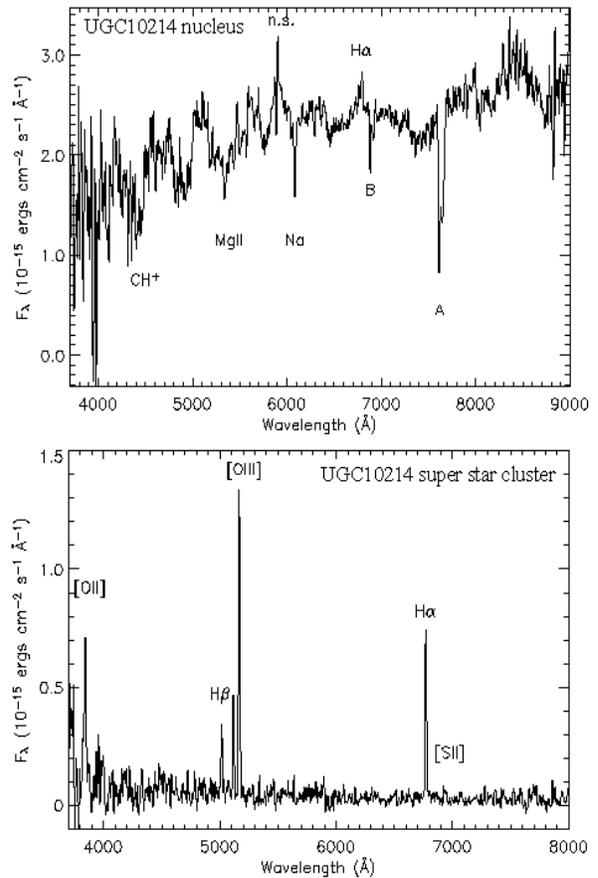

**Figure 5 -- Optical spectra of the UGC 10214 nucleus (top panel) and tidal tail massive star cluster, J160616.85+552640.6 (bottom panel). Residual telluric absorption bands are labeled A and B, while the 5892 Å sodium-D emission line is labeled "n.s.".**

Optical spectroscopy (§2.2) of a region spatially matched to the SED nuclear extraction, Figure 5 (top panel), reveals a relatively flat continuum with very weak Hα emission, atomic absorption MgII and Na bands, and molecular CH+ absorption. The derived redshift for these lines is 0.031, exactly matching the HI redshift measured by Briggs et al. (2001), establishing their origin with the Tadpole nucleus. The overall spectrum resembles that of late-type stars, ruling out the presence of a low-luminosity type-2 obscured AGN. We therefore have three lines of converging evidence: relatively blue stellar colors for the nuclear light, SED matching an S0-type galaxy and late-type stellar spectrum, all indicating that the nuclear region is powered by stellar light from mature stars. Evidently, the nucleus of the Tadpole remains in a quiescent state even after the galaxy-galaxy interaction. This is a surprising result given that UGC10214 is gas rich and therefore holds a reservoir capable of fueling a nuclear starburst (Figure 2). And indeed, many interacting spiral galaxies demonstrate active star formation in the nucleus and surrounding regions as gas is transferred inward with the loss of disk angular momentum from dissipative processes and non-axisymmetric forces. The Tadpole may be poised to actively form stars in the nucleus as the interaction matures, or alternatively, the gas and dust has been swept out of the nuclear region from the interaction. We explore this topic in more detail in section 4. Next, we consider the disk and extra-nuclear regions.

### 3.3 Disk and Extra-nuclear Region

Spectacular *HST* imaging (see §2.1) and spectroscopy of the Tadpole disk, VV29a, revealed a complex distorted bar-spiral-arm morphology in which the inner arm separates vertically (z-axis) from the disk plane, the outer spiral arm distorts horizontally, wrapping around the disk and unraveling out into the intergalactic medium. Both the tail and outer spiral arm are bright at optical wavelengths (including strong [OIII] and Balmer emission lines), as seen in Figure 1 and Figure 4 (top panel), indicative of early-type stars and recent, massive star formation. On the other hand, at optical wavelengths the inner arm, 'bar' and circum-nuclear regions appear more flocculent due to thick dust lanes and halo emission. This situation is remedied by the *Spitzer* observations, in which embedded star formation is revealed in the inner disk region: several bright 8 and 24 μm "hot spots" are seen against the arm and inter-arm regions forming a "ring". For illustrative purposes, we highlight three different regions, one along the eastern inner arm, the second along the western inter-arm and the third along the south-western inter-arm (note the labeled white arrows in Figure 4, top panel): J160605.90+552532.5, J 160602.28+552518.4, and J160601.48+552531.4, respectively.

The 12″ diameter aperture integrated fluxes for these regions are shown in Figure 4, bottom panel. The comparison SEDs are normalized to the 2.2 μm flux density of the first (eastern "hot spot") disk region (denoted with filled circles). As before, For clarity we only show the old stellar component at short wavelengths. The old stellar component requires a uniform foreground extinction of $A_v \sim 1$ mag applied to the "hot spot" fluxes to better match the model SEDs (see §3.2 for more details on this procedure). Note that the extinction is, in all probability, different for the other "hot spot" regions, but for figure clarity we do not show corresponding SED fits to the western "hot spots". At the shortest optical wavelengths, the Tadpole SED is significantly brighter relative to the "old" galaxy SED, consistent with a UV excess from ongoing massive star formation (§4.1). In the IR, the dust signatures are very strong, both in the dust continuum (24 μm) and PAH indicators (8 μm), characteristic of disk galaxies (e.g., Dale et al. 2005) and comparable to that of a normal Sc-type spiral galaxy[3]. Relative to the star light, the brightest dust signature comes from the "hot spot" in the southwestern inner-arm (SED denoted with magenta triangles), which is probably the most striking "red" feature seen in the disk region (Figure 3; Figure 4, top panel) owing to its location (between the inner and outer spiral arms) and strong 24 μm signal. Interestingly, this hot spot is proximally close, in projection at least, to the outer arm-to-tidal tail interface and the intruder galaxy VV 29c, estimated to be 100 kpc behind the Tadpole disk (Briggs et al. 2001). VV 29c is clearly seen in the *HST* images and in Figure 4 as the "blue" smudge along the western edge of the disk, directly between the *Spitzer* western hot spots. The companion is about 1/3 the (neutral hydrogen) mass of the main Tadpole disk, but is also gas rich and therefore should exhibit an infrared signature that contributes to the southwestern emission. In section 4.5 we examine the disk morphology in the context of a collisional ring galaxy interaction.

---

[3] Given that the integrated fluxes are coming from relatively large physical regions, each covering ~7.5 kpc diameter, which would contain numerous star formation regions, stellar clusters and diffuse ISM each with their own distinct IR properties, the detailed Tadpole measurements are closely related to global measurements.

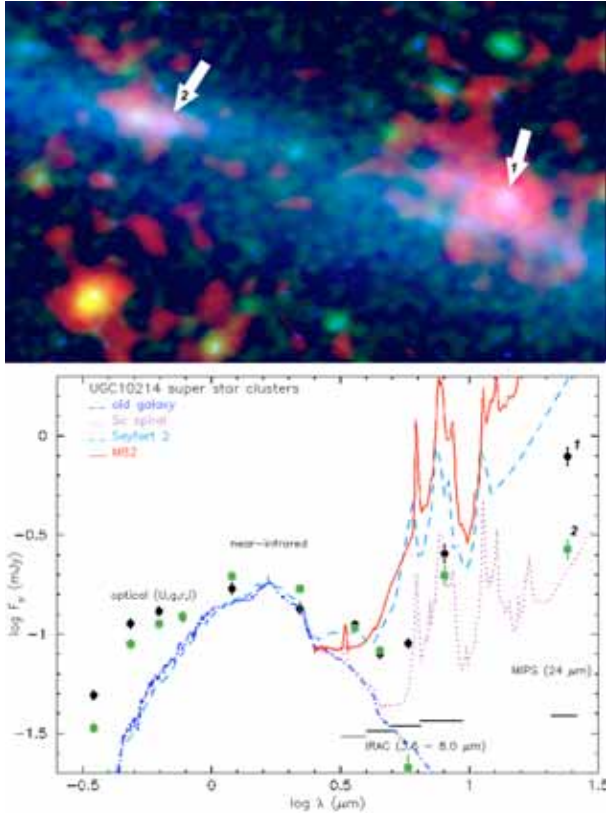

**Figure 6 -- Tidal tail clusters of the Tadpole Galaxy. The upper panel shows a 3-color RGB composite of the tail, where blue is g-band (F475W) from the *HST*-ACS (Tran et al. 2003), green is IRAC-1 (3.6 μm) and red is MIPS-1 (24 μm). The orientation is standard E of N. The x-axis field of view is ~1′. The lower panel shows the optical-infrared SED of the two super star clusters; denoted *1* (brighter 24 μm cluster to the west) and *2* (fainter 24 μm cluster to the east), integrated using a 12″ aperture. See Figure 4 for a description of the overlaid SEDs.**

### 3.4 Tidal Tail

The outer spiral arm of the Tadpole, VV29b, wraps northward and around the disk, behind or against (in projection) the inner spiral arm and intruder galaxy, exiting the disk to the south and extending another ~2′ (~75 kpc) eastward into the intergalactic medium. A schematic of the disk, spiral arms and tidal tail is given in Briggs et al. (2001). The plume's most striking feature is the filamentary string of young, massive stellar clusters that stretch from end to end and into the disk arms (e.g., De Grijs 2003; Tran et al. 2003). The most prominent cluster, or aggregate of clusters, is located about two-thirds of the tail length from the disk. At *HST* resolution, the "cluster" is resolved into a string of stellar associations, forming a "U-shaped" structure. The largest associations have masses large enough to qualify as globular clusters or even dwarf galaxies. These "super star clusters" have colors that are extremely blue, from strong Balmer and [OIII] emission lines (Figure 5), indicating that the light is overwhelmingly dominated by hot, massive stars[4]. Consequently, these clusters must be very young, a few mega-years, which is considerably offset from the dynamical age of the interaction, ~150 mega-years. The relative age difference is consistent with a scenario in which dynamical interaction and subsequent tidal distorting wave propagation is the trigger mechanism for formation of the star cluster. Although clusters are observed throughout the tidal tail, large gaps between clusters suggest that they are from staggered stochastic star formation, possibly self-gravitating structures. Whether they remain that way depends on their density and mass function.

IR observations provide a way to gauge the mass coming from older stars that may bolster the binding energy of the same region as the super star clusters. Here we focus on two such regions: (1) the most massive super star cluster, observed at 16h06m16.85s +55d26m40.6s, and (2) the second most prominent super star cluster, 16h06m20.2s +55d25m56.2s, located toward the eastern tail end. A view of these two super star clusters in given in Figure 6, top panel, which shows an RGB composite of the eastern tail region. The *HST*-ACS g-band image (in blue) is combined with the *Spitzer*-IRAC 3.6 μm image (green) and MIPS 24 μm image (red). The *HST* angular resolution has been preserved in order to discern the stellar clusters. The brightest super star cluster is located along the western edge of view, and the second, less prominent, cluster to the east.

---

[4] There is a remote, but provocative, alternative that the light is coming from a miniature version of an AGN; e.g., Farrah et al. 2005.

The 12″ diameter aperture integrated fluxes (§2.5) for these two regions are shown in Figure 6, bottom panel, and the channel luminosity for the brightest cluster is tabulated in Table 2. The comparison SEDs (which now include a Seyfert-2 template; §2.6) are normalized to the 2.2 μm flux density of the brightest super star cluster (filled circles). Although the set of SED templates, ranging from old stars to AGN/Seyfert class, do bracket the observations, it is clear that the SED templates are not adequate models of the star clusters, particularly at the short-wavelength regime where blue stars dominate the emission. More careful modeling of the cluster stars is needed to properly study the radiative transfer of the region. Consequently, we have not applied any extinction corrections to the super star cluster SEDs. In the optical, the extinction is probably low due to the massive stars disrupting and blowing away their birth clouds, consistent with the vivid, unobscured blue light seen in the optical imaging. A significant blue excess relative to the model templates is consistent with newly born massive stars or localized star formation. At long wavelengths, the IR signal is exceptionally strong relative to the NIR emission: the most prominent star cluster is also the brightest IR source in the tail. At the longest wavelength, the 70 μm signal probably arises from the same region as the 24 μm peak, but given the relatively poorer resolution, the cluster and tail environment (as well as local background galaxies) are not easily separated in the far-infrared (see Figure 2); Table 2 includes an upper limit measurement for the 70 μm flux density. We should stress that with the disparate spatial resolution between data sets, it is possible that the optical light and IR light are coming from different spatial regions -- the IR appears to peak on the northern side of the optical cluster structure, but coincident with the brightest emission in the HST-ACS images.

Optical spectroscopy of the super massive star cluster, J160616.85+552640.6, is shown in Figure 5 (bottom panel). This spectrum can be compared directly with that of Tran et al. (2003). The continuum is very weak, probably not detected by our observations. The most powerful emission line is [OIII], followed by the recombination lines Hα and Hβ, [OII] and [SII]. There is no detection of [NII] near the Hα line. The derived redshift is 0.031, matching the HI redshift of Briggs et al. (2001), establishing their Tadpole origin. The emission lines were fit with a gaussian profile and photometry extracted assuming the continuum was detected, or as is the case, the continuum was small relative to the nebular emission; Table 3 lists the resulting line fluxes. The low line ratios, including Balmer decrement[5] Hα/Hβ = 2.55, [OIII]/Hβ = 5.90, and [SII]/Hα=0.11, are consistent low metal abundances and high electron temperatures, not unlike metal-poor HII regions; e.g., compare with the line abundances of the HII regions in the low-metallicity SM-type galaxy NGC 55 (Otte & Dettmar 1999; Furgeson et al. 1996). The stellar cluster must represent the first epoch of star formation in Tadpole plume which is at least 100 Myr in age. In the next section we estimate the rate of star formation using the integrated line measurements.

The SED morphology and optical spectroscopy of the super star clusters can be summarized as: (1) strong excess at short wavelengths, from 0.3 to 1 μm, featuring strong emission lines in a low metallicity environment, dominating the total energy coming from the region (Table 2), (2) excess MIR emission between 3 and 5 μm, (3) strong PAH features from 3.3 to 11 μm, and (4) very strong warm dust continuum from ISM and young stellar objects. The blue excess (with strong emission lines) is consistent with massive star formation, whose SFR is discussed in §4.1, while the relative weakness of 2.2 μm continuum indicates that the super star clusters do not have a significant population of old stars. The young and hot stars are creating a strong UV radiation field that is transiently heating the surrounding small dust grains, creating a non-equilibrium emission in the short MIR window. The radiation field is also exciting the PAH molecules, which sit on top of a very steep continuum gradient that extends well beyond 24 μm. In the next section we further investigate the PAH features after removing the underlying dust continuum.

---

[5] Note that the Balmer decrement is not consistent with standard solar case-B recombination.

## 4 Discussion

In this section we consider the IR signatures of star formation, focusing on the nature of the ongoing production of stars in the disk and the newly created super massive clusters in the tidal tail.

### 4.1 Star Formation Rate

A variety of star formation diagnostics have been developed to quantify the rate at which massive stars are forming in galaxies (see Kennicutt 1998 for a review; Kewley et al. 2002; Condon 1992). Directly measuring the ionizing UV or recombination Hα emission from massive stars is feasible only when the foreground dust attenuation is low. Furthermore, calibration of the star formation rate (SFR) generally only applies to 'normal' metallicity disk galaxies with a Salpeter power-law initial mass function (IMF). Both of these limitations may translate into a factor of ~two uncertainty in the SFR for a range in galaxy types and metallicities (see Boselli et al. 2001). When the UV photons are absorbed by dust, reprocessing creates infrared emission that is proportional to the UV radiation field. Consequently, the appropriate SFR diagnostic for solar metallicity, gas-rich galaxies (including starbursts) employs the infrared signature of star formation.

The IR spectrum of galaxies is dominated by cold (T < 30 K) dust emitting in the FIR, and therefore optimally traces the massive star formation in dust-rich galaxies. Unfortunately, spatial resolution limitations of FIR observations (e.g., IRAS) restrict the SFR estimates of galaxies to global scales. Detailed SFR study is only possible at shorter wavelengths, in the MIR through ISO and *Spitzer* observations, for example. The source of MIR emission is complex, arising from starlight, the ionized ISM, cirrus (excited by starlight produced from lower mass stars), transiently heated dust and PAHs, thermal radiation from small dust grains, and non-thermal (synchrotron) radiation. Nonetheless, the MIR has been shown to be an effective tracer of massive star formation (e.g., Helou 2000; Förster Schreiber 2004; Helou et al. 2004; Calzetti et al. 2005). For the Tadpole Galaxy, we employ both the optical and MIR/FIR diagnostics to estimate the SFR in detail and in a global sense. The Tadpole is by no means 'normal', possessing distorted (interaction-shocked) spiral arms and a metal-poor tidal tail with massive stellar clusters likely possessing an extended IMF (see §4.4 below); consequently, we can expect sizeable scatter in the SFR estimates. The SFR analysis is nevertheless instructive toward contrasting the star formation in the nuclear, disk and tidal tail regions, as well as comparison with other interacting galaxies. As a final reminder, confidence in the SFR diagnostics rests on the assumption that the ionizing UV radiation is originating from young, massive stars, $M > 5\ M_o$. A non-zero contribution to the radiation field arises from the underlying stellar 'backbone' of relatively older, lower-mass stars, imparting a systematic overestimate in the SFR.

The global star formation rate for the Tadpole system may be estimated from the IRAS 60 and 100 μm emission, which was sufficiently sensitive to detect the cold dust in the disk, but not in the tidal tail. The IRAS FIR bands are used to estimate the total FIR between 40 to 120 μm, using the prescription of Helou et al. (1988):

$$F(FIR) = 1.26 \times 10^{-14} (2.58\ F60 + F100)\ [W\ m^{-2}], \quad (1)$$

where F60 and F100 are the IRAS flux densities at 60 and 100 μm in units of Jy. The corresponding FIR luminosity is $4\pi D_L^2 F(FIR)$, where $D_L$ is the luminosity distance. Following Kewley et al. (2002), the total IR, 8 to 1000 μm, star formation rate is:

$$SFR\ (IR)\ [M_o\ yr^{-1}] = 7.9 \times 10^{-44}\ L_{FIR}\ [erg\ s^{-1}]. \quad (2)$$

The IRAS 60 and 100 μm flux densities are 0.16 and 0.78 Jy, respectively, translating to a FIR flux of $1.50 \times 10^{-14}\ W\ m^{-2}$ and luminosity $3.01 \times 10^{43}\ erg\ s^{-1}$ assuming a distance of 129 Mpc to the Tadpole. We arrive at a SFR (IR) ~ 2.4 $M_o$ $yr^{-1}$. This value represents a lower limit to the total SFR since IRAS is not sensitive to the relatively faint star formation in the tidal tail.

The *Spitzer* 24 μm observations are, however, sensitive to the entire Tadpole system. We now utilize the MIR technique developed by Calzetti et al. (2005) in their detailed study of M51a using *Spitzer* and *HST*. In this work, the 24 μm observations of individual HII regions and complexes were calibrated against recombinant Pα (1.876 μm) observations, which proportionally measures the ionized radiation from the massive, forming stars in the spiral arms of M51a. They find a tight linear correlation, proving that (at least for M51a) the 24 μm dust continuum closely traces the star formation. The 24 μm and Pα $\nu L_\nu$ luminosities are related by (Calzetti et al. 2005):

$$\text{Log}\ [L_{P\alpha}] = 0.97\ \text{Log}\ [L_{24\mu m}] - 0.87, \quad (3)$$

where the luminosities have units of erg $s^{-1}$. The calibration should be reliable when the star formation density is high enough, > 1 $M_o$ $yr^{-1} kpc^{-2}$, to provide energy balance between the UV photons and the dust absorption. Using the global 24 μm luminosity measured for the Tadpole Galaxy (6.76 ×$10^{42}$ erg $s^{-1}$; Table 1), we deduce a corresponding Pα luminosity of 4.73 ×$10^{40}$ erg $s^{-1}$. The Pα luminosity is assumed to be directly related to the Hα luminosity under standard case B recombination: $F_{H\alpha} = 8.734 \times F_{P\alpha}$ (Osterbrock 1989), translating to an Hα luminosity of 4.13 ×$10^{41}$ erg $s^{-1}$. The SFR (Hα) is proportional to the total stellar luminosity emitted by young, massive stars (e.g., Kennicutt 1998):

$$\text{SFR (H}\alpha\text{)}\ [M_o\ yr^{-1}] = 7.9\times10^{-42}\ L_{H\alpha}\ [\text{erg s}^{-1}] \quad (4)$$

where SFR is the star formation rate in solar masses per year, assuming an IMF with a Salpeter slope and 'normal' galaxy metallicity. The global Hα luminosity implies a SFR equal to 3.26 $M_o$ $yr^{-1}$, agreeing to within 25% of the global value estimated from the FIR luminosity. The agreement is remarkably close considering that the MIR calibration comes from only one galaxy (M51a), which is unrelated in Hubble type and galaxy-galaxy interaction strength from that of the Tadpole. Infrared indicators of star formation may be remarkably robust with galaxy type and evolutionary stage. The *Spitzer* 70 μm observations (Table 1), like the IRAS 60 μm, effectively trace the current star formation in the Tadpole. Combining the *Spitzer* observations with the SFR calibrated for IRAS (e.g., Condon 1992; Cram et al. 1998; Hopkins et al. 2002), we compute a total 70 μm SFR (M > 0.1 $M_o$) of 4.5 $M_o$ $yr^{-1}$, consistent with the FIR estimate (Eq. 2) with inclusion of star formation from lower mass stars. Assuming the global SFR is ~2 to 4 $M_o$ $yr^{-1}$, the rate of massive star formation in the Tadpole is comparable to the global rate of late-type disk galaxies. From Figure 1, we know that the star formation is coming from distinct extra-nuclear sites in the disk and in the tidal tail, but not from the nucleus (§3.2). We now consider detailed SFR in the disk and tidal tail.

A large fraction of the Tadpole star formation is coming from dust-obscured complexes in the disk (see Figures 3-4). The hot spots have approximately the same 24 μm flux density; consider the eastern hot spot featured in Figure 4 (bottom panel; see also §3.3): J160605.90+552532.5. The 24 μm flux density is ~4 mJy, giving an infrared luminosity of ~9.9 ×$10^{41}$ erg $s^{-1}$. The corresponding Pα luminosity (Eq. 3) is 7.3×$10^{39}$ erg $s^{-1}$, which after conversion to the Hα luminosity leads to a SFR (Eq. 4) equal to ~0.5 $M_o$ $yr^{-1}$. This SFR is ~16% of the total SFR, representing a significant fraction of the total star formation in the Tadpole system. Assuming the three complexes featured in Figure 4 have the same SFR (a valid assumption given the comparable 24 μm flux densities), some 50% of the total star formation from massive stars is coming from the three most prominent infrared hotspots observed by *Spitzer*.

The optical spectral observations (Figure 5) provide integrated line fluxes (Table 3) that can be used to estimate the SFR in the tidal tail of the Tadpole Galaxy. For low extinction, the Hα line flux is the most reliable diagnostic, Eq. 4. Accordingly, the super star cluster, J160616.85+552640.6, Hα line flux (Table 3) converts to 0.23 $M_o$ yr$^{-1}$, a relatively high SFR density in comparison to other galaxies actively forming stars, including low metallicity blue compact dwarf galaxies (cf. Hopkins et al. 2002; Boissier et al. 2003). Another optical SFR indicator that has been shown to be effective is the integrated [OIII] + Hβ flux (Drozdovsky et al. 2005):

SFR ([OIII]+Hβ) [$M_o$ yr$^{-1}$]= 6×10$^{-42}$ $L_{[OIII]+H\beta}$ [erg s$^{-1}$] (5)

which gives a rate of 0.39 $M_o$ yr$^{-1}$ for the cluster after integrating the H$_\beta$ and two [OIII] lines fluxes in Table 3, or roughly a factor of ~1.7 larger than the SFR estimated from the Hα line. This diagnostic is more strongly dependent on the metallicity in comparison to the Hα SFR indicator. Moreover, both SFR estimates represent a lower limit since the extinction toward the cluster is non-zero, the (undetected) continuum is non-zero, and the low metallicity cluster environment is not typical of Milky Way star formation regions where the SFR relations are typically calibrated. Assuming the SFR is ~0.2 to 0.4 $M_o$ yr$^{-1}$, this single super star cluster represents 7 to 13% of the total star formation in the Tadpole system. In the MIR, the cluster exhibits a very strong 24 μm signal (Figure 6), indicative of hot dust heated by the strong UV radiation field associated with the massive stars. Since the physical conditions (including star formation density) in the tidal tail are quite dissimilar to those of M51a, it is very unlikely that the 24 μm to Pα correlation (Eq. 3) is valid for the super star cluster. And indeed, the resulting SFR (24 μm) is 0.11 $M_o$ yr$^{-1}$ according to the 24 μm luminosity, a factor of ~2 to 4 smaller than the estimate coming from the Balmer line emission (which is also likely to be too low). Furthermore, as noted in §3.4, if the infrared emission is arising from a source unrelated to the cluster, the star formation rates need to agree. The SFR estimates for the Tadpole system, disk and tidal tail are summarized in Table 4. We next consider the PAH emission bands as tracers of star formation.

## 4.2 Probing Star Formation with PAH Emission Diagnostics

MIR PAH emission is ubiquitously present throughout the Milky Way (Boulanger et al. 1998), while also central to the emission observed in nearby galaxies that are powered by star formation (Genzel et al. 1998; Lu et al. 2003). These broad emission bands arise from vibrational fluorescence of UV-excited PAH molecules in the diffuse ISM, and from within the photodissociation regions (PDRs) formed on the surfaces of neutral gas clouds exposed to the ionizing light of early-type stars. The most prominent bands are at 6.2, 7.7, and 11.2 μm, with weaker bands at 3.3 and 8.6 μm and at longer MIR wavelengths. These bands have been demonstrated to be effective tracers of ongoing star formation activity from early-type stars in dust-rich galaxies (e.g., Roussel et al. 2001; Forster Schreiber et al. 2004). The nearby galaxy M82, whose spectrum is shown in Figs 3 & 5, is a classic example of a dusty, actively forming stars galaxy with IR radiation dominated by PAHs at short wavelengths and a rising strong dust continuum at long wavelengths. The effectiveness of PAHs as a star formation indicator has its limits, however, as these molecules are destroyed by the intense radiation within HII regions and close to core AGN (cf. Genzel et al. 1998; Peeters, Spoon & Tielens 2004).

The *Spitzer* IRAC bands are sensitive to the PAH emission from low-redshift galaxies (e.g., Englebracht et al. 2005), although the short 3.6 μm and 4.5 μm channels are instead, to first order, largely indicators of photospheric starlight. Comparing the 5.8 and 8.0 μm channels with optical or NIR imaging produces a brilliant contrast between star formation regions and starlight from the older, underlying populations. Many beautiful examples exist, including *Spitzer* imaging of M81 (Wilner et al. 2004; Gordon et al. 2004), NGC 7331 (Regan et al. 2004; Smith et al. 2004), NGC 300 (Helou et al. 2004), M51 (Calzetti et al. 2005), and NGC4038/9 (Wang et al. 2004). The IRAC bands alone, for example comparing 3.6 μm with 8 μm, have been shown to be an effective discriminator between Hubble

Type morphology (Pahre et al. 2004b). With this context in place, we now interpret the IR signatures of the Tadpole Galaxy imaging.

For the Tadpole Galaxy, optical and IR images clearly show that ongoing star formation is present in the disk and tidal tail of the Tadpole. Embedded star formation is revealed in the IRAC 8 μm and MIPS 24 μm images, see Figures 1 - 6. The nucleus and disk IR SEDs are comprised of four components: starlight following a Rayleigh-Jeans decline toward long wavelengths, continua from transiently-heated small dust grains, PAH emission bands, most notably at 7.7 μm, and cooler dust emission that is rising beyond the 24 μm band. Stars appear to dominate the nuclear light (Figure 4, middle panel), whereas the PAHs and continuum are relatively weak or suppressed, which suggests that either the PAH bands are extincted by 10 μm silicate or ice absorption (highly unlikely given the lack of coincident 24 μm emission), or there is an overall lack of star formation in the nucleus. This is a remarkable finding, the system has matured ~100 Myr since the interaction and yet the nucleus is not undergoing a starburst or even modest star formation phase, very unusual for galaxy-galaxy systems (e.g., Xu et al. 2000). Intriguingly, the quiescent Tadpole nucleus is similar to that of the Antennae Galaxy nucleus (NGC4038/9), a spectacular merging system with deeply embedded massive star forming regions but whose nuclear light is dominated by old stars (Gilbert et al. 2000) even with the presence of molecular gas in the nucleus (Gao et al. 2001). For both the Tadpole and the Antennae the bulk of the star formation is occurring far from the nucleus in gas-rich density enhancements of the disk, triggered by the interaction between gas-rich spiral galaxies (see also §4.4 – 4.5).

The IR-bright regions of the Tadpole disk (Figure 4, bottom panel) indicate strong PAH and continuum emission, comparable to a late-type Sc galaxy, but not as strong (or continuum steep) as the starburst galaxies M82 (which includes silicate absorption) or NGC 7714 (no silicate absorption; Brandl et al. 2004). Well beyond the primary sites of star formation, the tidal tail super massive clusters (Figure 4) exhibit yet more bizarre behavior -- a large excess of starlight at short optical wavelengths, an excess of NIR light at 3 to 5 μm, very strong PAH bands riding atop a steep continuum that rises similar to that of M82, and a relatively high SFR density between 0.2 and 0.4 $M_o$ $yr^{-1}$ per cluster region. The current degree of activity in the disk is comparable to that of late-type spirals, while the massive clusters in the tidal tail have remarkable efficiency that are consistent with interaction-induced star formation. The dormant state of the nucleus may ultimately evolve toward a starburst phase as the dynamical interaction matures and the gas is locally concentrated, akin to the present state of the Antennae Galaxies (N4038/9). In the following section we discuss the state of the nucleus in more detail.

We now turn our attention to the IR emission arising from the super star clusters in the tidal tail. Interpretation of the IR emission bands is complicated by the underlying stellar light that dominates at short wavelengths, and the rising dust continuum that strengthens at long wavelengths. We attempt to decouple these components by removing the star light from all bands, and removing the dust continuum from the IRAC 8 μm channel, thus revealing the 7.7 μm PAH band. (Measuring the PAH strengths using only broad-band measurements is obviously fraught with difficulty, however the exercise is often useful toward better understanding of the IR emission arising from star formation throughout the galaxy, areas that are typically too large for spectral mapping.) For the starlight, the standard procedure is to scale and match the IRAC 3.6 μm channel to the expected starlight spectrum, removing this scaled component from the other IRAC channels (Helou et al. 2004). The major drawback to this technique is that the 3.6 μm channel is not pure starlight, but includes the 3.3 μm PAH band, as well as a hot dust component that arises from non-equilibrium UV photon absorption. Indeed, the Tadpole SEDs clearly show an excess in the 3.6 μm channel. A better alternative is to use the NIR 2 μm channel, which is less affected by dust extinction, as well as being the peak in the stellar component flux density. Accordingly, the NIR is used to normalize the "stellar" SED, here represented by the spectrum of an old elliptical galaxy (Silva et al. 1998) shown in Figure 4, which is then subtracted from the optical and IR flux densities. This procedure is only carried out on the integrated fluxes presented in the previous section. In principle, this may also be carried out on a pixel-by-pixel basis, thus creating finer-scale images, but in practice this is complicated by the necessity to apply aperture and calibration corrections to the photometry, which are critically sensitive to scale sizes (as discussed in the previous sections).

For the case of dust continuum removal from the 8 μm photometry, we make the simple assumption that the 4.5, 5.8 and 24 μm channels bracket and define the underlying dust continuum. This is not strictly the case at 5.8 μm, due to the 6.2 μm PAH band that overlaps with this channel. Even the IRAC 4.5 μm band may contain residual PAH emission, which blanket the entire MIR window from 3 to 15 μm. Consequently, we can expect uncertainties that range between 10 to 20%, including systematically over-estimating the continuum (conversely, underestimating the PAH strength). Due consideration of the 5.8 μm bandpass with respect to the PAH width, redshifted to the Tadpole radial velocity, coupled with the relative strength of the 6.2 μm to the 7.7 μm PAH bands, we estimate that the PAH component of the 5.8 μm channel is roughly 10% the strength of the 7.7 μm PAH component. Using this prescription, we compute the slope of the dust continuum, anchored between 4.5 and 24 μm, remove it from the 8 μm flux, thereby isolating the PAH emission at 7.7 μm, and likewise, separating the 6.2 μm PAH component from IRAC 5.8 μm channel. This procedure is iterated until the PAH components and underlying continuum converge to a solution. The resulting flux density should represent the PAH emission band at 7.7 μm. Figure 7 shows the SED for the super massive star clusters after removal of the starlight and the dust continuum from the 8 μm channel. The MIR continuum is represented by the dashed (black) line.

Figure 7 showcases the remarkable properties of the super massive star clusters in the tail. At optical wavelengths, the excess (relative to an old stellar population) is attributed to young, massive stars that are tightly clustered into (possibly) self-gravitating systems. These hot stars radiate most of their energy in the ultra-violet, but within the window studied here most of the energy is coming from the g-band (0.5 μm) channel (e.g., Table 2), which is consistent with the *HST*-ACS observations. In the NIR, the 3.6 μm channel shows a clear excess, possibly due to the 3.3 μm PAH band in addition to a hot dust continuum.

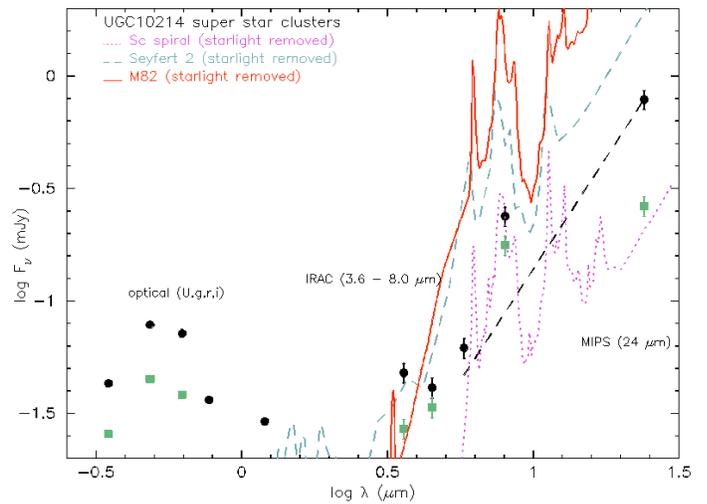

**Figure 7** -- **SEDs of the tidal tail clusters, with the underlying starlight and dust continuum removed from the IRAC 8 μm band. This diagram represents optical and infrared light that is in excess to the underlying old stellar population (Fig 3). The dashed line represents the hot dust continuum, stretching from 5.6 to 24 μm, and is removed from IRAC-4 (8 μm) to create the integrated 7.7 μm PAH line emission (see §4.2). See Figure 4 for a description of the overlaid SEDs.**

The longer IRAC channels are under the influence of a steeply rising dust continuum that extends well beyond 24 μm. The 8 μm channel, corrected to represent only the 7.7 μm PAH band, exhibits a strength that is comparable or greater than that of the dusty spiral SED template. Finally, the 24 μm channel is much brighter than would be expected for quiescent spiral galaxy star formation, but instead suggests a heating mechanism from massive star formation. We deduce that the super massive star cluster has formed within a dense molecular cloud association that either survived the gravitational interaction with subsequent fragmentation or formed after the collision created density enhancements in the distorted spiral arm that would become the tidal tail (see also the analysis of Wallin 1990). The cloud later collapsed in a violent burst of star formation. The high star formation rate, 0.23 $M_o$ yr$^{-1}$, deduced from the optical nebular lines and the MIR dust emission is comparable to that seen in other massive clusters in interacting galaxies (e.g., Stephan's Quintet; Lisenfeld et al. 2004) and dwarf galaxies (Braine et al. 2001; Annibali et al. 2003). The luminous stars have blown a hole in their birth cloud, revealing the nascent cluster. The powerful radiation field, coupled with a large reservoir of gas and dust (Figure 2), is producing a strong IR signal that resembles nuclear starbursts, yet is remarkably far from the actual nucleus of the Tadpole. Under this scenario, however, we may expect the 7.7 µm PAH bands to be suppressed as the UV-radiation field could be hard enough to destroy these molecules.

To assess the strength of the PAH emission, we compare its integrated flux with both the underlying 8 µm continuum (similar to a technique used by Genzel et al. 1998) and the 24 µm continuum. The ratio of the 8 µm PAH flux density to the underlying dust continuum flux density for the Tadpole reveals a distribution that ranges from a low of 0.34 in the nucleus (where PAHs are weak), to a high of ~2 corresponding to an embedded star formation region in the disk (Figure 4, bottom panel). The disk PAH strength is comparable (to a factor of ~two) to the strength of the 7.7 µm PAH line in starbursts and ULIRGS (Genzel et al. 1998). For the super massive star clusters in the tidal tail, the ratio is ~1.6 for the brightest cluster (Figure 7), relatively strong in comparison to the nucleus and quiescent regions of the disk, but smaller than the disk star formation regions. A similar result is found for the nuclear super star cluster in the nearby dwarf galaxy NGC 1705 (Cannon et al. 2005). The PAH band strength for the clusters is weaker in comparison to the disk star formation, presumably due to their destruction. But it is interesting to note that the 4.5 and 5.8 µm channels show a strong IR excess, which may be due to the relatively weaker 6.2 µm PAH band emission, which in some cases represents a more robust PAH indicator (e.g., Peeters, Spoon & Tielens 2004).

### 4.3 Origin of the IR Radiation: Quiescent or Active Star Formation?

Comparing the PAH strength with the 24 µm emission, which is purely due to hot thermal dust radiation, provides contrast between quiescent and active star formation. Preliminary studies from *Spitzer* observations seem to indicate that the $F_{8\mu m} / F_{24\mu m}$ ratio is, for the most part, constant for global measurements of normal disk galaxies (cf. Helou et al. 2004). At finer scales the ratio is seen to decrease -- the PAH strength weakens

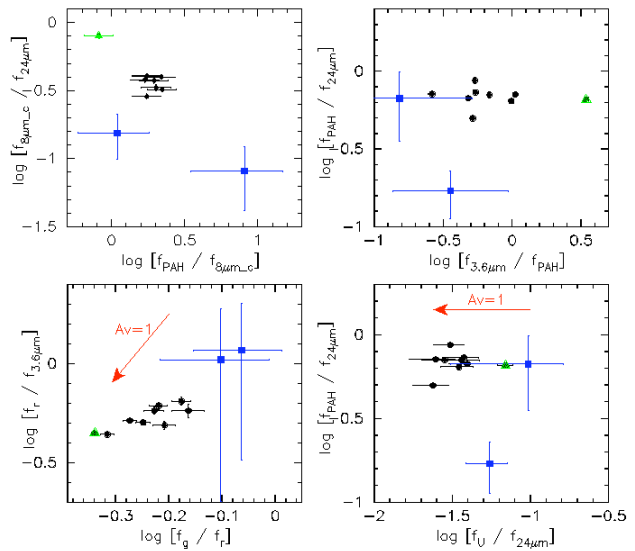

**Figure 8** -- **Mid-infrared colors of the Tadpole Galaxy. The upper two panels show the log [$F_{8\mu m}/ F_{24\mu m}$] ratio versus 8 µm flux density, but with the right panel representing the 7.7 µm PAH line (in which starlight and dust continuum is removed from the 8 µm channel). The lower two panels show the MIR color versus the IRAC-1 (3.6 µm) flux density, which represents the underlying stellar mass distribution. The right panel, as before, represents the 7.7 µm PAH line. Filled circles indicate the disk locations, and filled squares the super star clusters in the tidal tail. The nucleus is the green filled triangle. The dashed line represents the mean trend seen in the color-mag diagrams**

relative to the hot dust continuum emission -- when the star formation activity and the FIR continuum increases (Calzetti et al. 2005; as in circum-nuclear regions of starbursts, see Smith et al. 2004). The Tadpole results are shown in Figure 8, where we plot the log [$F_{8\mu m}/F_{24\mu m}$] color versus the near/mid-IR fluxes for eight regions in the disk (including the nucleus) and two tidal tail regions (Figure 6). As with the SED measurements, the photometry comes from 12″ diameter circular apertures, corrected for calibration and aperture effects.

The color vs. flux diagrams demonstrate the effect of removing both the underlying stellar light and dust continuum from the 8 μm channel, revealing the 7.7 μm PAH band. The left panels show the color with no modification, so it represents a combination of PAH, continuum and starlight. The figure shows a clear trend with brightness, where the ratio decreases with correspondingly decreasing integrated flux. But after correcting the 8 μm channel for starlight and continuum effects, the trend disappears; see right panels in Figure 8. Here the nucleus, disk star formation regions and the fainter super star cluster all have about the same color ratio (within the measurement uncertainty), ranging between -0.3 and -0.1 in the log. The one glaring exception is the prominent super massive star cluster, which has a ratio of -0.8 in the log, which is significantly smaller than any other location in the Tadpole. This is consistent with the SED results (Figure 7): the 24 μm emission is comparatively strong, while the 7.7 μm PAH indicator is suppressed in this region, presumably due to grain destruction. The ratio is not only much smaller relative to the Tadpole disk, but is also much smaller than what is observed in nuclei and star formation regions from nearby spiral galaxies (based on the preliminary results of SINGS; Murphy et al. 2005).

The equivalent width of the PAH emission, or comparing the emission line strength to the underlying MIR continuum, can be used to discriminate between dust heating and IR emission components, including AGN, starbursts, PDR and normal HII regions (Genzel et al. 1998; Laurent et al. 2000; Förster Schreiber et al. 2004; Peeters, Spoon & Tielens 2004). Normalizing the 7.7 μm PAH strength (§4.2) with the continuum derived in the IRAC 8 μm band, we derive the equivalent width diagnostic for the Tadpole Galaxy. The results are shown in Figure 9 (upper left panel) for the nucleus, disk hot spots and tidal tail super star clusters, where we have plotted the PAH diagnostic versus the 8 to 24 μm dust continuum color. In this color plane, disk star formation is separated from the nuclear emission and the tidal tail star clusters. The nucleus is virtually absent of emission from PAH or dust continuum at 24 μm, graced by only continuum at 8 μm that arises from starlight in the Rayleigh-Jeans tail. The star formation hot spots in the disk cluster around a similar continuum color and PAH equivalent width, indicative of their common formation conditions. Employing the diagnostic analysis from Laurent et al. (2000), the disk hot spots have a PAH equivalent width, ~2, and 24 to 8 μm dust continuum color, ~3, that is consistent with a combination of HII and PDR origin. In the tidal tail, the bright 24 μm super star cluster exhibits a very weak PAH signature, relative to the MIR 'hot' dust continuum, both indicative of PAH destruction from the hard UV radiation field produced by the massive star cluster. In comparison to the Laurent et al. (2000) diagnostic curves, the PAH equivalent width, ~1, and continuum color, ~6, of the super star cluster is not consistent with HII or PDR emission, but more closely resembles that of AGN -- namely, hot dust and PAH destruction from a powerful UV radiation field. Toward the end of the tidal tail the fainter star cluster's 8 μm emission is almost exclusively PAH dominated. Its equivalent width, ~8, and continuum color ~13, are only consistent with HII origin.

Optical-to-infrared colors are sensitive to differences in stellar population, extinction and ISM components. Comparing optical g'-band with Ks-band or IRAC 3.6 μm traces extinction and stellar components, which should correlate with dust emission. Figure 3 compares the dust-extincted optical emission with the star formation sensitive 8 μm emission. Quantifying the extinction, we compute the g'-3.6 μm color (flux ratio $F_{3.6\mu m}/F_{g'}$) and compare with the 8 μm emission after subtraction of the underlying stellar component (as prescribed by Helou et al. 2004). The color ranges from 2 to 8.5, where the larger values correspond to dust lanes seen in Figure 3. The color ratio is measuring the differential extinction between 0.5 and 3.6 µm, highest near the sites of recent star formation in the extranuclear disk arms (compare also with Figure 3). Assuming the optical-IR color is solely due to dust reddening and that the late-type stellar population is the same throughout the disk, the maximum extinction corresponds to $A_v$ ~3.4 mag, representing the mean extinction across the resolution element of the Spitzer observations (FWHM ~ 2.5″, or 1.5 kpc). In reality the color represents a combination of extinction and old versus intermediate stellar components. The stellar populations are investigated in Figure 9, plotting optical-to-infrared colors. The [$F_g$/ $F_r$] color compares B-type stars to older main-sequence stars, while [$F_r$ / $F_{3.6\mu m}$] color compares main-sequence to the underlying stellar backbone of evolved stars. The $F_U$ to $F_{24\mu m}$ flux ratio should be sensitive to the embedded star formation since the U-band is tracing the unobscured massive stars and 24 µm emission the dust that is heated by the UV-radiation produced by these stars. The reddening due to selective extinction, important for the disk star formation regions, only affects the optical and NIR colors (i.e., horizontal shift in the diagram). The main results are: (1) the reddest colors are coming from the nucleus and circum-nuclear regions, (2) the bluest colors correspond to the tidal tail clusters, and (3) the color-color trends are indicative of a combination of reddening due to selective extinction and stellar population -- the youngest, most massive stellar populations are found in the tidal tails, while the evolved populations are found in the Tadpole nucleus and inter-arm regions. The range in color exceeds the amount expected simply due to dust extinction (see Figs 4, 6), indicative of a large span in stellar type. The colors and PAH equivalent width diagnostic of the super massive cluster, J160616.85+552640.6, are significantly different from the corresponding disk and nuclear values. The IR-bright super massive star cluster, J160616.85+552640.6, is unique in both optical and MIR windows. At the extreme end of the tidal tail, the fainter super star cluster, J160620.2s+552556.2, has colors and PAH strengths that are more comparable to the star formation regions in the disk (see also Figure 7, which shows a similar trend), which may indicate that tidal tail star clusters have fundamentally different environmental conditions from beginning to end.

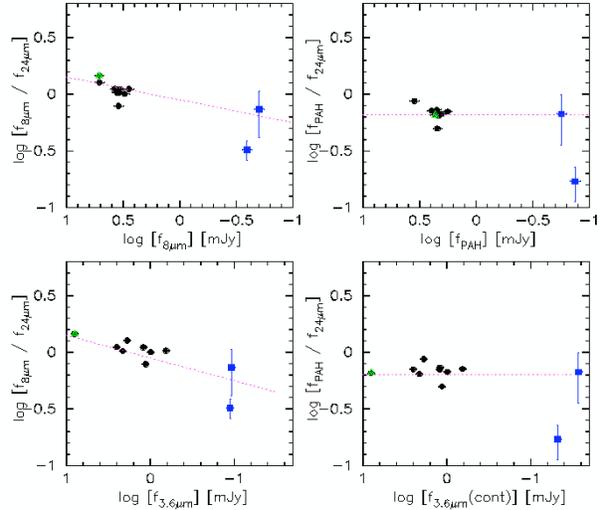

**Figure 9** -- Optical-infrared color-color diagrams. The upper left panel shows the 8 µm to 24 µm dust continuum ratio versus the 7.7 µm PAH line to the 8 µm dust continuum. The PAH band strength is estimated by removing the starlight and dust continuum from the 8 µm measurements. The upper right panel shows the infrared $F_{PAH}$ to $F_{24\mu m}$ color versus the infrared $F_{3.6\mu m}$ to $F_{PAH}$ color. The lower left panel shows the r'-band to infrared 3.6 µm ratio versus the g'-band to r'-band ratio. The lower right panel shows the optical U-band to infrared 24 µm color to infrared $F_{PAH}$ to $F_{24\mu m}$ color. The green triangle denotes the nucleus; filled circles indicate the disk locations, and filled squares the super star clusters in the tidal tail. The reddening vector is indicated in both panels.

## 4.4 Is the Tadpole Poised for a Nuclear Starburst?

Redistribution of gas from the outer to the inner parts of galaxies is a complex process that remains poorly understood, although several compelling mechanisms are under spirited test. Removal of angular momentum from spiral arms through gas dissipation and dynamical disruption should create inward flowing gas and thereby inducing a nuclear starburst, as Toomre and Toomre (1972) first conjectured for galaxy-galaxy interactions. Indeed, disrupted galaxies tend to exhibit enhanced star formation and nuclear activity (e.g., Soifer et al. 1984; Kennicutt et al. 1987; for a counter example, see Bergvall, Laurikainen & Aalto 2003). And yet the nucleus of UGC10214 shows little sign of activity as we infer from the lack of H$\alpha$ and mid-infrared (dust) emission (see Figs 4-5), even in the presence of significant neutral gas in the nucleus and disks of the primary (VV29a) and the secondary (VV29c) galaxies (Fig 2). Do we interpret this null result as the absence of gas flowing from the disk to the core as both a source of fuel and shock excitation to compress and fragment the gas? Our Spitzer observations seem to rule out the presence of a large scale bar of gas and dust, thus denying the means for non-axisymmetric funneling of gas from the disk to the core.

Like the gas-rich, bar-free Antennae system (NGC4038/9), the Tadpole nucleus is conspicuously quiet, possessing an old stellar population that dominates the nuclear light (see also Bergvall, Laurikainen & Aalto 2003 for other examples of quiescent nuclei of interacting galaxies). Are the star formation properties of the disk and nucleus of the Tadpole indicative of the type of galaxy-galaxy interaction, lack of a large scale bar, or due to the early stage of the encounter? As inferred from the tidal tail morphology and stellar cluster ages, the interaction is estimated to be 100 to 200 Myr in the past (Briggs et al. 2001; de Grijs et al. 2003), still relatively early in the merger process. The Tadpole tail morphology is similar to other tidal tail systems, which have been studied with detailed n-body simulations showing that the tidal tails tend to hold together and coalesce back into the larger system as the merger evolves over giga-year time scales (e.g., Hibbard & Mihos 1995; Struck & Smith 2003; see next section). The current state of the nucleus may be transitory as the system evolves dynamically from a young interaction to a maturing merger-like system (assuming the companions are bound), accreting plumes of gas, dust and stars that have evolved beyond the parent disk, slowly transforming a 'dormant' nucleus into a gas-rich, active star-forming core. The time scale for such a transformation would be ~hundreds of mega-years as the companion and parent galaxies perform their slow gravitational dance, dissipating angular momentum through dynamical friction and large-scale "bar" creation. The process is complex, however, as a nuclear starburst is not only dependent on a large supply of fuel, but also a triggering mechanism to induce gravitational collapse in the interstellar gas. The merging Antennae Galaxies have sufficient levels of gas to fuel a nuclear starburst, but evidently the cold gas has yet to collapse in a burst of star formation, suggesting the lack of a trigger mechanism.

The core and circum-nuclear region of the Tadpole may yet hold secrets that could be revealed with detailed observations of the molecular gas (as traced by CO emission, for example) and high energy emission (e.g., UV and x-ray radiation). We have initiated a program to observe the Tadpole with mm-wave observations using the IRAM 30-m telescope and with UV observations from GALEX. A future paper will report the results of these recently acquired observations. Let us now consider quite another evolutionary scenario that concerns the rare class of interactions that create "ringed" galaxies. Below we re-examine the Tadpole morphology.

## 4.5 Off-Center Ring-Producing Collision?

In a number of ways the Tadpole exhibits the classic signature of a spiral-spiral interaction as plausibly described by the work of Toomre and Toomre (1972). Most notably, the unwrapping of spiral arms (tidal tails), plumes and counter-tails, are clearly evident in the radio observations of Briggs et al. (2001) and the detailed optical imaging from HST. The infrared imaging presented in this work, on the other hand, hints at yet even more fascinating Tadpole morphology. In sharp contrast to the nuclear region, the disk is vigorously forming new and massive stars. But unlike the optical imaging, the near/mid-infrared morphology appears as a "ring" of star formation; see Figs 3-4. This active star formation is hidden at optical wavelengths by the veiling column of dust in the disk and arms of the Tadpole, but is betrayed by the longer-wavelength infrared cameras of the Spitzer Observatory.

Is this ring of star formation a real, coherent structure? If so, we may hypothesis that the "ring" was produced or at least, enhanced by the interaction. This would not be a classic ring that arises from a direct face-on collision between two nearly-equal mass galaxies, but instead the more likely off-center collision that creates asymmetric rings of dynamically compressed stars and gas in the presence of spiral arms (see the excellent review by Struck 1999). The companion galaxy (VV29c) is smaller than the primary (VV29a), but is still relatively close in mass (~1/3 the total neutral hydrogen mass; Briggs et al. 2001). A companion of this relative size and an off-nuclear collision would induce substantial tidal torques that spawn radial-expanding compression waves. The resulting "ring" structure would be traced by the higher density of stars and the newly forming stars embedded within their birth clouds. The near-infrared (tracing older stars) and the mid-infrared (tracing warm dust from massive stars) imaging presented in Figure 1 paints a compelling (if qualitative) argument for dynamically-induced ringed star formation: ring-like structure of both older stars and newly forming stars, outer spiral arms (mostly quiescent), distorted tidal tail, and a companion that is seen directly behind the primary it its moderately inclined orientation to the line of sight. It is therefore plausible that the Tadpole system represents a spiral galaxy closely colliding with a larger spiral galaxy some 100 to 200 Myr ago. Dynamical rings have another peculiar property that is relevant to this discussion. They tend to have very little nuclear star formation. Although the models predict gas infall due to a rebounding inward traveling wave, in actuality ringed galaxies usually have quiescent nuclei. The head-on or off-center collision must somehow sweep gas away from the nuclear region, leaving the nucleus dormant or extinct of star formation (see also discussion in Appleton, P. N., Charmandaris, V., & Struck, C. 1996). The Tadpole's primary galaxy nucleus may be quiescent because gas was propelled away from the center region into the outer disk region where it was later compressed into molecular clouds which collapsed to form new stars. Not all of the gas is stripped from the nucleus, but the close interaction may remove enough to arrest cloud collapse.

A proper examination of this ring hypothesis would entail n-body, hydro-dynamical simulations of the system. To date the Tadpole has not been fully modeled. It is instructive to note that the Tadpole has morphological similarities with another two systems that has been analyzed with detailed computer simulations. Struck & Smith (2003) modeled in depth the ring system NGC 7714/14 (Arp 284) and Smith et al (2005) the system Arp 107. Like the Tadpole, Arp 284 has a massive, gas-rich disk with distorted arms and tidal tail, plumes and counter-tails, and a prominent ring-like arm structure in the disk. Their simulations show that the off-center collision with a relatively lower-mass companion provokes the observed large-scale distortions, flinging gas away from the outer spiral arms and creating tidal plumes and bridges, while also inducing a shock-induced ring of star formation. It all happens within 100 to 200 Myr after the interaction, similar to what is observed with the Tadpole interaction time scale. The intriguing similarities between these two systems should also be balanced with the differences. Arp 284 has a nuclear starburst (see e.g., Brandl 2004), and, owing to its much closer proximity to the Sun, possesses an overall morphology that is more complex than that of the Tadpole (admittedly the Tadpole may possess similar complexities, such as bridges of gas and stars between the companions, but due to the distance and orientation to the line of sight we may never know). Nevertheless, it would seem a promising route to carry out full n-body, hydro-dynamical simulations of the Tadpole system to gain quantitative understanding of the interaction history and future evolution of the underlying stellar and ISM components.

## 4.6 Tidal Tail Stellar Cluster Mass

The nature and evolution of the super massive star clusters associated with dynamically interacting galaxies is of great interest to astronomers. It has been suggested that these clusters represent emerging globular cluster systems, or self-gravitating dwarf galaxies (akin to primordial versions of the Omega Centauri object orbiting the Milky Way); see de Grijs (2003) for a review. Questions remain, however, as to the evolution of these objects -- do they remain bound clusters or disperse into open clusters (see van den Bergh 1995; Kroupa 1998)? The key parameters are the total mass and size of the clusters, as well as the environment gas pressures, which determine whether the systems will survive future tidal disruptions and dynamical evaporation. The distribution of the mass itself, the shape of the initial mass function, is crucial toward deducing the total mass of the system. The presence of massive stars is no guarantee that lower mass stars, which dominate the cumulative numbers observed in Galactic clusters and globular clusters, are also present. Given the complexity of measuring the IMF, the usual assumption is that the mass distribution follows the standard relation (e.g., Salpeter function). A top-heavy IMF would render mass estimates to be systematically too large. With this caveat in mind, we press forward with estimating the cluster mass.

Tran et al. (2003) investigated the properties of the Tadpole tidal clusters, including the IR-bright source investigated in this study (see Table 2). They deduce a total mass ~0.5 to $1 \times 10^6$ $M_o$ in a small region of ~0.1 kpc$^2$ (half-light region), which implies a very high gas pressure and star formation efficiency to overcome the destructive behavior of O stars at the end of their short lives (see Elmegreen & Efremov 1997). However, compared to globular clusters in the Milky Way, the Tadpole SSC is spread over a larger area, diluting the mass density and rendering the cluster vulnerable to disruption. This may thus be an example of an super star cluster that does not remain bound over the lifetime of the interaction. Nevertheless, a number of SSCs in a variety of galaxies have properties that are promising examples of newly emerging galaxies and globulars (see Bastian et al. 2005; Lisenfeld et al. 2004; Braine et al. 2001; Pasquali et al. 2003; Gilbert et al. 2000).

The optical-IR measurements presented in this work can be used to estimate the cluster mass in a region constrained by the data beam size, but are unfortunately not sufficient in angular resolution to address the mass density issue. We will carry out two separate calculations, the first using our "standard" aperture (§2.5) employed throughout this study, and the second using a smaller aperture that is better optimized to higher resolution imaging from our near-infrared observations. Starting with g-band (0.5 μm) photometry of the IR-bright cluster, J160616.85+552640.6, the luminosity is found to be ~35 × 10$^7$ $L_o$ in a 12″ aperture (Table 2). The M/L ratio for young forming (single burst) stars is significantly different from that measured in the field due to the luminous phase of young clusters (e.g., generally flatter IMF, reflecting the dominance of massive stars; see also de Grijs 2003; Figer et al. 1999). Chandar et al. (1999) compile evolutionary models that indicate the M/L in the visual band is ~0.01 to 0.02 for a 10$^{6.7}$ year burst. The corresponding stellar mass of the cluster is ~5 × 10$^6$ $M_o$ based on the optical luminosity. This g'-band luminosity, however, includes a significant contribution from oxygen and hydrogen emission lines.

A better diagnostic comes from the NIR, 2 μm, which directly traces the underlying starlight. Here we measure a luminosity of ~9 × 10$^7$ $L_o$ in a 12″ aperture (Table 2). To estimate the M/L ratio, we combine the evolutionary models of Leitherer & Heckman (1995) showing V-K colors for young forming clusters (we adopt the solar metallicity solutions) with the Chandar et al. (1999) and Battinelli & Capuzzo-Dolcetta (1989) V-band evolutionary models, to construct a 2 μm M/L ~ 0.05 to 0.1. Although a large uncertainty in the M/L ratio is unavoidable given the strong reliance upon the models inputs (e.g., Salpeter IMF), we note that this estimate is roughly consistent with K-band studies of young clusters in nearby galaxies (e.g., Knapen et al. 1995; Wada et al. 1998; Ryder & Knapen 1999; Vallejo, Braine Baudry 2002). The corresponding stellar mass follows as ~7 × 10$^6$ $M_o$, which, as expected, is larger than the optical light estimate of Tran et al. (2003). Since the ground-based optical and K-band images have much better angular resolution than the *Spitzer* data, we now deploy an aperture that is better matched to the angular size of the super star cluster. The imaging data limit the minimum aperture size to 4.5" (~2.8 kpc) in diameter. In the g'-band, the integrated flux is 0.045 mJy, corresponding to a luminosity of $13.9 \times 10^7$ $L_o$, and a mass of $\sim 1.4 \times 10^6$ $M_o$. At 2.2 μm, the integrated flux is 0.033 mJy, or a luminosity of $2.3 \times 10^7$ $L_o$, giving a stellar mass of $\sim 1.6 \times 10^6$ $M_o$, or about a factor of two larger than the Tran et al. (2003) derivation.

This large mass is, however, balanced by the lower mass density that follows from the larger area studied here: $\sim 2.5 \times 10^5$ $M_o$ kpc$^{-2}$ in surface density. The Tadpole super star cluster is much larger in area than the typical super star cluster found in nearby galaxies (e.g., O'Connell, Gallagher, & Hunter 1994), and is therefore vulnerable to dynamical disruption unless the interaction shock pressure remains high enough to hold the cluster together (a scenario discussed by Elmegreen & Efremov 1997). On the other hand the region is rich with multiple clusters, forming 'chain'-like aggregates as seen in the high resolution HST imaging (Fig 6, e.g.). The evolution of such long-lived objects may lead to classic globular clusters and dwarf satellite galaxies (see for example the work of Kroupa 1998; Fellhauer & Kroupa 2002).

Further work remains to understand the high-mass stellar populations, however, as the massive O and B stars, emit most of their radiation in the ultra-violet window (shortward of the SWIRE U-band). Their IMF properties and evolution may be revealed at these energetic wavelengths. Recent studies of tidal tail star formation using GALEX UV-imaging (cf. Neff et al. 2005) look very promising toward deciphering the nature of super massive star formation in the tidal tails of interacting galaxies.

## 5   Summary

Owing to its fortuitous location in the ELAIS N1 region, the *Spitzer* Wide-area Infrared Extragalactic Survey (SWIRE) observed the interacting galaxy UGC10214 with deep ground-based optical and NIR imaging in support of the *Spitzer* IR imaging, compiling a comprehensive data set ideally suited to study of the current generation of star formation in this remarkable system. In this work we presented color-composite imaging of the "Tadpole" that spans from 0.3 to 70 μm. We considered both the young stellar component, underlying stellar mass distribution, hot dust continuum and PAH emission from the Tadpole ISM and star formation regions, focusing on the nucleus, extra-nuclear regions of the disk, spiral arms, and the tidal tail structures. The major results of this study are as follows:

- the IR morphology is broadly separated by: nucleus, disk with star formation "hot spots", spiral arms, plume and tidal tail structures; there is no evidence for a large-scale bar
- active star formation is observed in the disk, spiral arms, and tidal tail, with an estimated cumulative SFR (IR) equal to ~2 to 4 $M_o$ yr$^{-1}$
- the nucleus is powered by starlight from the old stellar population; there is no evidence of active star formation; the deduced average foreground extinction is <1 $A_v$
- IR-bright "hot spots" in the disk exhibit strong PAH emission bands, tracing the locations where gas has merged to form massive star formation regions; the level of star formation activity is comparable to that of late-type spiral galaxies, with an estimate SFR ~ 0.5 $M_o$ yr$^{-1}$ in the eastern cluster J160605.90+552532.5, representing ~17% of the total star formation in the Tadpole; the combined disk SFR (MIR) is estimated to be >50% of the total in the system.
- the inner spiral arm or ring-like structure exhibits bright IR emission, indicative of embedded massive star formation; the "ring" may be the result of an off-center collision with a lower-mass companion galaxy
- the outer spiral arm is very blue, dominated by starlight from intermediate age and massive stars; the arm distorts into a long

- tidal tail that extends 2′ (75 kpc) into the intergalactic medium
- the tidal tail is very blue, dominated by starlight, lined with super massive star clusters; two of these clusters are detected with *Spitzer* in the MIR -- the Tadpole is an example of an off-nuclear or tidal-tail starburst
- the low metallicity, IR-bright tidal tail super star cluster, J160616.85+552640.6, exhibits many remarkable IR properties, including exceptionally strong 24 μm emission relative to all other optical-IR components; PAH emission is suppressed, suggesting grain destruction from the hard UV radiation field; the SFR is estimated to be 0.2 and 0.4 $M_o$ yr$^{-1}$, or 10% of the total in the Tadpole system
- the mass of the super star cluster is estimated to be $1.4 - 1.6 \times 10^6$ $M_o$ based on the g'-band (0.5 μm) and NIR (2.2 μm) flux measured in a 4.5″ aperture centered on the cluster; the mass is comparable to the largest globular cluster found in the Milky Way and to the so-called "tidal dwarf galaxies".

Three outstanding results from this study are: (1) even though the system has matured ~100 to 200 Myr since the interaction, the nucleus is not undergoing a starburst or even modest star formation phase, unusual for galaxy-galaxy systems, (2) the main disk is actively forming stars in a distorted spiral arm or "ring"-like structure, and (3) within the tidal tail, far from the main disk, a massive cluster of newly formed stars appears to have formed only a few Myr ago from a large reservoir of metal-poor gas and dust detected by the *Spitzer* Space Telescope.

*Acknowledgements*
We would like to thank Phil Appleton for tapping his formidable knowledge of collisional ring galaxies, and to Helene Roussel, Igor Drozdovsky, Seppo Laine, Steve Lord, Joe Mazzarella and Bernhard Schultz for many spirited discussions on starburst galaxies. A special thanks to several members of the SINGS team, including George Helou, J.D. Smith, Lee Armus, and Eric Murphy, for imparting their insightful knowledge of infrared galaxies and *Spitzer* data reductions. Finally, we would like to express our gratitude to the anonymous reviewer of this paper for many helpful suggestions that greatly strengthened the interpretations. The *Spitzer* Space Telescope is operated by the Jet Propulsion Laboratory, California Institute of Technology, under contract with NASA. This research has made use of the NASA Extragalactic Database, operated by IPAC/Caltech, under contract with NASA.

## List of Tables

**TABLE 1** -- UGC10214 Global Flux Density and Luminosity per Filter

| Band | $\lambda_o$ (μm) | $\Delta\lambda$ (μm) | $F_\nu$ (mJy)[a] | $\nu L_\nu$ ($10^9 L_o$)[b] |
|---|---|---|---|---|
| U | 0.3580 | 0.0638 | 2.132 | 8.89 |
| g' | 0.4846 | 0.1285 | 9.035 | 27.8 |
| r' | 0.6240 | 0.1347 | 15.292 | 36.6 |
| i' | 0.7743 | 0.1519 | 19.848 | 38.3 |
| J | 1.250 | 0.162 | 47.181 | 56.4 |
| Ks | 2.150 | 0.397 | 43.020 | 29.9 |
| IRAC-1 | 3.600 | 0.750 | 23.674 | 10.2 |
| IRAC-2 | 4.500 | 1.020 | 13.816 | 4.58 |
| IRAC-3 | 5.800 | 1.414 | 15.762 | 4.06 |
| IRAC-4 | 8.000 | 2.974 | 28.253 | 5.27 |
| MIPS-24 | 24.00 | 5.35 | 27.093 | 1.69 |
| MIPS-70 | 71 | 19 | 207.5 | 4.36 |

Notes: (a) no reddening corrections are applied here; flux uncertainties are ~5% for the optical and NIR measurements, ~10 to 20% for the IRAC and MIPS measurements; (b) assuming a distance to UGC10214 of 129 Mpc.

**TABLE 2** -- Flux Density and Luminosity per band for super cluster: J160616.85+552640.6

| Band | $\lambda_o$ (μm) | $\Delta\lambda$ (μm) | $F_\nu$ (mJy)[a] | $\nu L_\nu$ ($10^7 L_o$)[b] |
|---|---|---|---|---|
| U | 0.3580 | 0.0638 | 0.049 | 20.4 |
| g' | 0.4846 | 0.1285 | 0.113 | 34.8 |
| r' | 0.6240 | 0.1347 | 0.131 | 31.4 |
| i' | 0.7743 | 0.1519 | 0.122 | 23.5 |
| J | 1.250 | 0.162 | 0.170 | 20.3 |
| Ks | 2.150 | 0.397 | 0.135 | 9.38 |
| IRAC-1 | 3.600 | 0.750 | 0.112 | 4.65 |
| IRAC-2 | 4.500 | 1.020 | 0.0800 | 2.65 |
| IRAC-3 | 5.800 | 1.414 | 0.0900 | 2.32 |
| IRAC-4 | 8.000 | 2.974 | 0.255 | 4.76 |
| MIPS-24 | 24.00 | 5.35 | 0.789 | 4.91 |
| MIPS-70[c] | 71 | 19 | <0.13 | <0.27 |

Notes: (a) 12″ circular aperture; flux uncertainties are ~5% for the optical, 10% for the NIR measurements, ~10 to 20% for the IRAC and MIPS measurements; (b) assuming a distance to UGC10214 of 129 Mpc; (c) 70 μm flux deduced from the mean surface brightness over a 20″ aperture.

**TABLE 3** -- Optical integrated line fluxes for super massive cluster: J160616.85+552640.6

| Line (Å) | $\lambda_o$ (Å) | flux ($10^{-15}$ erg/cm²/s) | FWHM (Å) |
|---|---|---|---|
| [OII] 3737 | 3840.8 | 18.9 ± 0.7 | 21.5 |
| Hβ 4861 | 5015.2 | 5.7 ± 0.3 | 11.6 |
| [OIII] 4959 | 5115.6 | 0.8 ± 0.1 | 11.1 |
| [OIII] 5007 | 5165.3 | 25.8 ± 1.0 | 12.0 |
| Hα 6563 | 6770.2 | 14.6 ± 0.5 | 12.4 |
| [SII] 6718 | 6925.7 | 0.10 ± 0.01 | 12.4 |
| [SII] 6733 | 6944.2 | 0.07 ± 0.01 | 12.4 |

**TABLE 4** -- Current Star Formation Rates

| SFR Indicator[a] | Global SFR[b] ($M_o$ yr$^{-1}$) | Disk Hotspot[c] ($M_o$ yr$^{-1}$) | Super Star Cluster[d] ($M_o$ yr$^{-1}$) |
|---|---|---|---|
| FIR[e] | 2.37 | -- | -- |
| MIPS 70 μm[f] | 4.5 | -- | -- |
| MIPS 24 μm | 3.26 | 0.51 | 0.11 |
| Hα | -- | -- | 0.23 |
| [OIII]+H$_\beta$ | -- | -- | 0.39 |

Notes: (a) $M > 5 M_o$; (b) total area = 2.97 sq. arcmin (see §3.1); (c) J160605.90+552532.5; (d) J160616.85+552640.6; (e) using IRAS 60 and 100 μm emission; (f) derived for $M > 0.1 M_o$.